\pdfoutput=1
\documentclass[aps,superscriptaddress,amsmath,amssymb,nofootinbib,secnumarabic,a4paper]{revtex4} 

\usepackage{graphicx}% Include figure files

\allowdisplaybreaks[3]
\newcommand{\ti}[1]{#1}  %keep titles, etc, of refs
\newcommand{\vol}[1]{\bf #1}  %journal volume style

\newcommand{\flow}[1]{\mathcal{#1}}
\newcommand{\vect}[1]{\boldsymbol {#1}}
\newcommand{\tens}[1]{\boldsymbol{\mathsf{#1}}}
\newcommand{\vecti}{\boldsymbol{{I}}}

\newcommand{\tscale}{\theta}
\newcommand{\Vscale}{\mathcal{V}}

\newcommand{\G}{{\mathbf g}}
\newcommand{\GAMMA}{\pmb{\gamma}}
\newcommand{\stepindex}{\upsilon}
\newcommand{\I}{\mathbf {I}}
\newcommand{\compress}{\kappa}

\begin{document}

\title[Jaynes' MaxEnt, Riemannian Metrics and Least Action Bound]{Jaynes' Maximum Entropy Principle, Riemannian Metrics and Generalised Least Action Bound}% Force line breaks
\author{Robert K. Niven}
\email{r.niven@adfa.edu.au}
\affiliation{School of Engineering and Information Technology, The University of New South Wales at ADFA, Canberra, ACT, 2600, Australia.}
\author{Bjarne Andresen}
\affiliation{Niels Bohr Institute, University of Copenhagen, DK-2100 Copenhagen \O, Denmark.}
%\email{r.niven@adfa.edu.au}
%end aps

\date{2 July 2009; corrected 24 January 2011}% can use \today

\begin{abstract}
The set of solutions inferred by the generic maximum entropy (MaxEnt) or maximum relative entropy (MaxREnt) principles of Jaynes -- considered as a function of the moment constraints or their conjugate Lagrangian multipliers -- is endowed with a Riemannian geometric description, based on the second differential tensor of the entropy or its Legendre transform (negative Massieu function). The analysis provides a generalised {\it least action bound} applicable to all Jaynesian systems, which provides a lower bound to the cost (in generic entropy units) of a transition between inferred positions along a specified path, at specified rates of change of the control parameters. The analysis therefore extends the concepts of ``finite time thermodynamics'' to the generic Jaynes domain, providing a link between purely static (stationary) inferred positions of a system, and dynamic transitions between these positions (as a function of time or some other coordinate). If the path is unspecified, the analysis gives an absolute lower bound for the cost of the transition, corresponding to the geodesic of the Riemannian hypersurface.  The analysis is applied to (i) an equilibrium thermodynamic system subject to mean internal energy and volume constraints, and (ii) a flow system at steady state, subject to constraints on the mean heat, mass and momentum fluxes and chemical reaction rates.  The first example recovers the {\it minimum entropy cost} of a transition between equilibrium positions, a widely used result of finite-time thermodynamics. The second example leads to a new {\it minimum entropy production principle}, for the cost of a transition between steady state positions of a flow system.

\end{abstract}
\maketitle

%% %%############################################################################
\section{\label{sect:intro}Introduction} 
%% ############################################################################
%

Jaynes' {\it maximum entropy principle} (MaxEnt) and its extension, the {\it maximum relative entropy principle} (MaxREnt), based on the principles of inductive (probabilistic) rather than deductive reasoning, arguably constitutes one of the most important tools for the solution of indeterminate problems of all kinds \cite{Jaynes_1957, Jaynes_1963, Tribus_1961a, Tribus_1961b, Shore_J_1980, Kapur_K_1992, Jaynes_2003}. In this method, one maximises the entropy function of a system -- a measure of its statistical spread over its parameter space -- subject to the set of constraints on the system, to determine its ``least informative'' or ``most probable'' probability distribution \cite{Jaynes_1957, Jaynes_1963, Jaynes_2003}. By a series of generic ``Jaynes relations'', this can then be used to calculate the macroscopic properties of the system, providing the best (inferred) description of the system, subject to all that is known about the system. Since its inception half a century ago, the MaxEnt and MaxREnt principles have been successfully applied to the analysis of a diverse range of systems, including in thermodynamics (its first and foremost application), solid and fluid mechanics, mathematical biology, transport systems, networks, economic, social and human systems \cite{Jaynes_1957, Jaynes_1963, Tribus_1961a, Tribus_1961b, Shore_J_1980, Kapur_K_1992, Jaynes_2003, Kapur_1989b, Kapur_K_1987}. 

The aim of this study is to examine a valuable extension to Jaynes' generic approach, by endowing the set of solutions inferred by Jaynes' method -- considered as a function of the set of moment constraints and/or their conjugate Lagrangian multipliers -- with a Riemannian geometric interpretation, using a metric tensor furnished directly by Jaynes' method. The analysis leads to a generalised {\it least action bound} applicable to all Jaynesian systems, which provides a lower bound for the cost (in generic entropy units) of a transition between different inferred positions of the system.  The analysis therefore extends the concepts of ``finite time thermodynamics'', developed over the past three decades \cite{Weinhold_1975a, Weinhold_1975b, Weinhold_1975c, Weinhold_1975d, Weinhold_1976, Salamon_A_G_B_1980, Salamon_N_A_B_1980, Salamon_B_1983, Salamon_I_B_1983, Andresen_1983_book, Salamon_N_I_1984, Andresen_B_O_S_1984, Salamon_N_B_1985, Schlogl_1985, Feldman_etal_1985, Nulton_etal_1985, Feldman_etal_1986, Levine_1986, Nulton_S_1988, Salamon_etal_1988, Andresen_etal_1988a, Andresen_etal_1988b, Hoffmann_etal_1989, Andresen_G_1994, Salamon_N_1998, Salamon_etal_2001, Schaller_etal_2001, Beretta_1986, Beretta_1987, Beretta_2008, Diosi_etal_1996, Crooks_2007, Feng_C_2009, Brody_H_2009}, to the generic Jaynes domain. The analysis reveals a deep, underlying connection between the essentially static manifold of stationary positions predicted by Jaynes' method, and lower bounds for the cost of dynamic transitions between these positions. %To the authors' knowledge, this generic connection has not been expounded previously.

The manuscript proceeds as follows.  In \S\ref{Jaynes}, the theoretical principles of Jaynes' MaxEnt and MaxREnt methods are discussed, followed by an appraisal of a generalised free energy (generalised potential) concept associated with Jaynes' method. In \S\ref{Riemann}, the concepts of a Riemannian metric, arc length and action sums and integrals are developed in a generic Jaynesian context, leading to a generic least action bound for transitions on the manifold of Jaynes solutions. Considerations of minimum path lengths, involving calculation of the geodesic in Riemannian space, are also discussed. In \S\ref{Apps},  the foregoing principles are applied to (i) an equilibrium thermodynamic system subject to mean internal energy and volume constraints, and (ii) a flow system at steady state, subject to constraints on the mean heat, mass and momentum fluxes and chemical reaction rates.  The first example (\S\ref{Equil}) recovers the {\it minimum entropy cost} of a transition between equilibrium positions, a widely used result of finite-time thermodynamics. The second example (\S\ref{Flow}) leads to a new {\it minimum entropy production principle}, for the cost of a transition between steady state positions of a flow system. The analyses reveal the tremendous utility of Jaynes' MaxEnt and MinXEnt methods augmented by the least action bound, for the analysis of probabilistic systems of all kinds.

%############################################################################
%
\section{\label{Jaynes}Jaynes' Generic Formulation (MaxREnt)} 
\subsection{Theoretical Principles}

The usefulness of Jaynes' method for statistical inference arises from its {\it generic} formulation, first expounded by Jaynes and other workers in the context of information theory \cite{Jaynes_1957, Jaynes_1963, Tribus_1961a, Tribus_1961b, Shore_J_1980, Kapur_K_1992, Jaynes_2003}, but which can be reinterpreted using a combinatorial framework (the ``Boltzmann principle'')\index{Boltzmann principle} \cite{Boltzmann_1877, Planck_1901, Vincze_1972, Grendar_G_2001, Niven_2005, Niven_2006, Niven_G_2009, Niven_2009_EPJB}. In consequence, the method can be applied to {\it any} probabilistic system involving the allocation of entities to categories; this includes -- but is not restricted to -- thermodynamic systems. For maximum generality, it is useful to include source or prior probabilities\index{prior probabilities} $q_i$ associated with each category $i=1,...,s$, to give the {maximum relative entropy} (MaxREnt)\index{maximum relative entropy principle (MaxREnt)} or {minimum cross-entropy} (MinXEnt) principle\index{minimum cross-entropy principle (MinXEnt)}. In the event of equal $q_i$, this reduces to the special case of Jaynes' {maximum entropy} (MaxEnt) principle\index{maximum entropy principle (MaxEnt)}\index{Jaynes!maximum entropy principle (MaxEnt)} \cite{Jaynes_1957, Jaynes_1963, Tribus_1961a, Tribus_1961b, Shore_J_1980, Kapur_K_1992, Jaynes_2003}. 

The MaxREnt method proceeds as follows.  To infer the ``least informative''\index{least informative distribution} or ``most probable"\index{most probable distribution} distribution of a probabilistic system, we wish to identify its observable {\it realization} or {\it macrostate} of maximum probability $\mathbb{P}$.  This is equivalent to maximising the following dimensionless function, chosen for several ``nice'' mathematical properties \cite{Boltzmann_1877, Planck_1901}:
\begin{equation}
\mathfrak{H} = \frac{1}{N} \ln \mathbb{P},  
\label{eq:Boltzmann1}
\end{equation}
For a system of $N$ distinguishable entities allocated to $s$ distinguishable categories, it can be shown that the distribution is governed by the multinomial distribution\index{multinomial distribution} $\mathbb{P} = N! \prod\nolimits_{i=1}^s q_i^{n_i} / n_i!$, where $n_i$ is the occupancy of the $i$th category and $N=\sum\nolimits_{i=1}^s n_i$. In the asymptotic limit $N \to \infty$, \eqref{eq:Boltzmann1} reduces to the relative entropy function\index{relative entropy function} \cite{Jaynes_1963} (the negative of the Kullback-Leibler function\index{Kullback-Leibler function} \cite{Kullback_L_1951, Kullback_1959}):
\begin{equation}
\mathfrak{H} = -\sum\limits_{i = 1}^s {p_i \ln \frac{{p_i }}{{q_i }}} 
\label{eq:relH}
\end{equation}
where $p_i=n_i/N$ is the frequency or probability of occupancy of the $i$th category. Maximisation of \eqref{eq:relH} is subject to the normalisation constraint \index{constraint!normalisation} and any moment constraints\index{constraint!moment} on the system:
\begin{gather}
\sum\limits_{i = 1}^s {p_i }= 1,
\label{eq:C0} 
\\
\sum\limits_{i = 1}^s {p_i f_{ri}}= \langle {f_r } \rangle, \quad r = 1,...,R, 
\label{eq:Cr}
\end{gather}
where $f_{ri}$ is the value of the property $f_r$ in the $i$th category and $\langle {f_r} \rangle$ is the mathematical expectation of $f_{ri}$.  Applying Lagrange's method of undetermined multipliers\index{Lagrangian method of undetermined multipliers}\index{calculus of variations} to (\ref{eq:relH})-(\ref{eq:Cr}) gives the stationary or ``most probable'' distribution of the system\index{stationary position of a system}\index{most probable position of a system} (denoted *):
\begin{gather}
\begin{split}
p_i^* &=  q_i e^ {  - \lambda_0   - \sum\limits_{r = 1}^R  {\lambda_r  f_{ri}}    } = \frac {1}{Z}  q_i e^ {  - \sum\limits_{r = 1}^R  \lambda_r  f_{ri}   }  , \\
%\label{eq:pstar1_i} \\
Z &=  e^{\lambda_0} = \sum\limits_{i=1}^s {q_i e^ {  - \sum\limits_{r = 1}^R  \lambda_r  f_{ri}   }} 
%\label{eq:Z_q}
\end{split}
\label{eq:pstar2_i}
\end{gather}
where $\lambda_r$ is the Lagrangian multiplier\index{Lagrangian multiplier} associated with the $r$th constraint, $Z$ is the partition function\index{partition function} and $\lambda_0=\ln Z$ is the Massieu function\index{Massieu function} \cite{Tribus_1961b}. In thermodynamics, the constraints $\langle f_r \rangle$ are usually taken to represent conserved quantities, and thus correspond to extensive variables\index{extensive variables} (e.g.\ internal energy, volume and numbers of particles), whilst the multipliers $\lambda_r$ emerge as functions of the intensive variables\index{intensive variables} of the system (e.g.\ temperature, pressure and chemical potentials).  It is useful to preserve this distinction between extensive and intensive variables, even beyond a thermodynamic context.

By subsequent analyses \cite{Jaynes_1957, Jaynes_1963, Tribus_1961a, Tribus_1961b, Shore_J_1980, Kapur_K_1992, Jaynes_2003, Niven_CIT}, one can derive the maximum relative entropy $\mathfrak{H}^*$ and the derivatives of $\mathfrak{H}^*$ and $\lambda_0$ for the system\index{Jaynes!MaxEnt relations}:
\begin{gather}
{ \mathfrak{H}^*  = \lambda_0  +  \sum\limits_{r = 1}^R \lambda_r \langle {f_r} \rangle }
\label{eq:Hstar}
\\
{\frac{{\partial \mathfrak{H}^*}}{{\partial \langle {f_r } \rangle }} = \lambda_r }  
\label{eq:MaxREnt_deriv1}
\\
{\frac{{\partial ^2 \mathfrak{H}^*}}{{\partial \langle {f_m } \rangle \partial \langle {f_r } \rangle }} 
= \frac{{\partial \lambda_r }}{{\partial \langle {f_m } \rangle }} = - g_{mr} \in - \G
\label{eq:MaxREnt_deriv2}}
\\
\frac{\partial  \lambda_0 }{\partial \lambda_r}  = - \langle {f_r} \rangle 
\label{eq:Massieu_deriv1}
\\
\begin{split}
\frac{\partial^2 \lambda_0}{\partial \lambda_m \partial \lambda_r}   = \langle {{f_r}{f_m}} \rangle- {\langle {f_r} \rangle}{\langle {f_m} \rangle}   
= - \frac {\partial \langle {f_r} \rangle}{\partial \lambda_m}  
= \gamma_{mr} \in {\GAMMA} 
\label{eq:Massieu_deriv2}
\end{split} 
\end{gather}
The second derivatives of $\lambda_0$ in \eqref{eq:Massieu_deriv2} express the dependence of each constraint on each multiplier, and therefore give the ``capacities'' or ``susceptibilities''\index{capacities}\index{susceptibilities} of the system (e.g.\ in thermodynamics, they define the heat capacity, compressibility, coefficient of thermal expansion and other material properties \cite{Callen_1960, Callen_1985, Weinhold_1975c}). Their matrix $\GAMMA$, the variance-covariance matrix of the constraints, is equal to the inverse of the matrix $\G$ of second derivatives of $\mathfrak{H}^*$ in \eqref{eq:MaxREnt_deriv2} (with change of sign), yielding the generic Legendre transformation\index{Legendre transformation} between the $\mathfrak{H}^*(\langle f_1 \rangle, \langle f_2 \rangle, ...)$ and $\lambda_0(\lambda_1, \lambda_2, ...)$ descriptions of the system \cite{Jaynes_1963}:
\begin{equation}
\G \, \GAMMA = \I,
\label{eq:Legendre}
\end{equation}
where $\I$ is the identity matrix \cite{Jaynes_1963}.  From \eqref{eq:MaxREnt_deriv2} or \eqref{eq:Massieu_deriv2} and the equality of mixed derivatives, we also obtain the generic reciprocal relations\index{Jaynes!reciprocal relations} $ {\partial \langle {f_r} \rangle}/{\partial \lambda_m} = {\partial \langle {f_m} \rangle}/{\partial \lambda_r}$ for the system. 

Jaynes also showed that the incremental change in the relative entropy can be expressed as \cite{Jaynes_1957}:
\begin{gather} 
{d\mathfrak{H}^* = \sum\limits_{r = 1}^R {\lambda_r \Bigl( d\langle {f_r } \rangle  - \langle {df_r } \rangle \Bigr)}  = \sum\limits_{r = 1}^R {\lambda_r \delta Q_r }}
\label{eq:dHstar}
\end{gather}
where $\delta W_r = \langle {df_r } \rangle = \sum\nolimits_{i=1}^s  p_i^* d f_{ri}$ and $\delta Q_r = \sum\nolimits_{i=1}^s  d p_i^* f_{ri}$ can be identified, respectively,  as the increments of ``generalised work''\index{Jaynes!generalised work} and ``generalised heat''\index{Jaynes!generalised heat} associated with a change in the $r$th constraint, and $\delta (\cdot)$ indicates a path-dependent differential.  Eq.\ \eqref{eq:dHstar} gives a ``generalised Clausius equality''\index{Jaynes!generalised Clausius equality} \cite{Clausius_1865}, applicable to all multinomial systems in the asymptotic limit.

It is again emphasised that the above relations \eqref{eq:pstar2_i}-\eqref{eq:dHstar} apply to {\it any} probabilistic system of multinomial form, in the asymptotic limit.  Although originally derived in thermodynamics, the above-mentioned quantities need not be interpreted as thermodynamic constructs, but have far broader application. Furthermore, the relations  \eqref{eq:pstar2_i}-\eqref{eq:dHstar} apply to the stationary position of any multinomial probabilistic system. The derivatives \eqref{eq:MaxREnt_deriv1}-\eqref{eq:Massieu_deriv2} therefore relate to transitions of the system between different stationary positions, or in other words, to paths on the manifold of stationary positions\index{manifold!of stationary positions}. Whilst the lack of inclusion of non-stationary positions may seem unnecessarily restrictive, such geometry provides a sufficient foundation for most of engineering and chemical equilibrium thermodynamics.  As will be shown, it is also useful for the analysis of many other systems of similar probabilistic structure.

\subsection{\label{Gen_Free_Energy}The Generalised Free Energy Concept} 

It is instructive to insert \eqref{eq:dHstar} into the differential of \eqref{eq:Hstar} and rearrange in the form:
%\newpage
\begin{align}
\begin{split}
d \phi &= - d \lambda_0  =  -d \ln Z =  \sum\limits_{r = 1}^R {\lambda _r  \delta W_r }  +  \sum\limits_{r = 1}^R d \lambda_r \langle {f_r} \rangle
%\\&
=-d\mathfrak{H}^*  +  \sum\limits_{r = 1}^R \lambda_r d \langle {f_r} \rangle  +  \sum\limits_{r = 1}^R d \lambda_r \langle {f_r} \rangle 
\end{split}
\label{eq:Massieu}
\end{align}
The negative Massieu function $-\lambda_0$ is therefore equivalent to a potential function $\phi$ which captures all possible changes in the system, whether they be in the entropy, constraints or multipliers.  For constant multipliers, it simplifies to the weighted sum of generalised work on the system. It thus provides a dimensionless analogue of the free energy concept used in thermodynamics.  For constant multipliers, $\phi \rvert_{\{\lambda_r\}}$ also provides a measure of the dimensionless ``availability''\index{availability}, or the available ``weighted generalised work''\index{Jaynes!generalised work}, which can be extracted from a system.  By extension of the principles of equilibrium thermodynamics, we can thus adopt the potential $\phi$ as a measure of distance from the stationary state\index{Jaynes!Massieu function}\index{Jaynes!generalised potential}. The system will converge towards a position of minimum $\phi$, representing the balance between maximisation of entropy within the system $\mathfrak{H}^*$, and maximisation of the entropy generated and exported to the rest of the universe by the transfer of generalised heats $\delta Q_r$ (see \cite{Niven_MEP} for further discussion). The advantage of Jaynes' generic formulation is that $\phi$ can be defined for {\it any} multinomial probabilistic system, and is not restricted to thermodynamic systems \cite{Jaynes_1957, Jaynes_1963, Jaynes_2003, Tribus_1961a, Tribus_1961b}. 

Returning to the second derivatives in the last section, we see that $\lambda_0$ can be replaced by $-\phi$ in \eqref{eq:Massieu_deriv1}-\eqref{eq:Massieu_deriv2}. The latter provides a clean (doubly negative) Legendre transformation\index{Legendre transformation} between matrices $\G$ and $\GAMMA$, and thus between the $\mathfrak{H}^*(\langle f_1 \rangle, \langle f_2 \rangle, ...)$ and $\phi(\lambda_1, \lambda_2, ...)$ representations of a system.  Furthermore, since $\GAMMA$ is equal to the variance-covariance matrix of the multipliers \eqref{eq:Massieu_deriv2}, it is positive definite (or semi-definite if singularities exist) \cite{Kapur_K_1992}. Since $\G$ is the inverse of a positive definite matrix \eqref{eq:Legendre}, it also is positive definite (or semi-definite) \cite{Kapur_K_1992}. The signs of $\G$ and $\GAMMA$, as defined in \eqref{eq:MaxREnt_deriv2} and \eqref{eq:Massieu_deriv2}, were chosen consistent with positive rather than negative definiteness, for reasons which will become clear in the next section.

\section{\label{Riemann} Riemannian Geometric Concepts}
\subsection{\label{Metric}Generalised Riemannian Metrics and Arc Lengths} 

Since the time of Gibbs \cite{Gibbs_1873b, Gibbs_1875}, examination of the geometry of the manifold of stationary positions has been of tremendous interest to scientists and engineers. In thermodynamics, this has typically involved analysis of the concave hypersurface defined by the Euler relation $S(\tilde{X}_1, \tilde{X}_2, ...)$, where $S$ is the thermodynamic entropy and $\tilde{X}_r$ are the extensive variables, or alternatively of its Legendre transform, the convex hypersurface $\psi(Y_1, Y_2, ...)$ or $F(Y_1, Y_2, ...)$, where $\psi=F/T$ is a Planck potential function, $F$ is a free energy and $Y_r$ are the intensive variables \cite{Callen_1960, Callen_1985}. Such interpretations have led to major advances in the understanding and analysis of thermodynamic processes and cycles \cite{Gibbs_1873b, Gibbs_1875}. However, adoption of the Jaynes MaxEnt framework (\S\ref{Jaynes}) permits a rather different insight, based on a Riemannian geometric interpretation.  As will be evident from the previous discussion, this interpretation extends well beyond ``mere'' thermodynamics, forming a natural adjunct of Jaynes' generic formulation (\S\ref{Jaynes}).  

Consider the $R$-dimensional hypersurface parameterised by the constraints $\{ \langle f_r \rangle \}$ or their conjugate Lagrangian multipliers $\{ \lambda_r \}$, representing the hypersurface of stationary states within the $(R+1)$-dimensional space given by $(\mathfrak{H}^*, \{ \langle f_r \rangle \})$ or $(\phi, \{\lambda_r \})$.  Since $\G$ \eqref{eq:MaxREnt_deriv2} and $\GAMMA$ \eqref{eq:Massieu_deriv2} are positive definite (i.e.\ $\vect{x}^\top \G \vect{x}>0$ or $\vect{x}^\top \GAMMA \vect{x}>0$ for any non-zero vector $\vect{x}$ \cite{Kreysig_1991, Trasarti_Battistoni_2002, Zwillinger_2003}), they can be adopted as Riemannian metric tensors\index{Riemannian metric} associated with the stationary state hypersurface defined by $\{ \langle f_r \rangle \}$ or $\{ \lambda_r \}$, and used to interpret its geometric properties. Indeed, even if $\G$ or $\GAMMA$ are positive semidefinite due to the occurrence of singularities (i.e.\ $\vect{x}^\top \G \vect{x} \ge 0$ or $\vect{x}^\top \GAMMA \vect{x} \ge 0$ for $\vect{x} \ne \vect{0}$), they can still be adopted as pseudo-Riemannian metric tensors on the stationary hypersurface.  This representation was first proposed by Weinhold \cite{Weinhold_1975a, Weinhold_1975b, Weinhold_1975c, Weinhold_1975d, Weinhold_1976}, and its implications in terms of a least action bound were subsequently developed, largely within a thermodynamic context, by Salamon, Berry, Andresen, Nulton and co-workers \cite{Salamon_A_G_B_1980, Salamon_N_A_B_1980, Salamon_B_1983, Salamon_I_B_1983, Andresen_1983_book, Salamon_N_I_1984, Andresen_B_O_S_1984, Salamon_N_B_1985, Schlogl_1985, Feldman_etal_1985, Nulton_etal_1985, Feldman_etal_1986, Levine_1986, Nulton_S_1988, Salamon_etal_1988, Andresen_etal_1988a, Andresen_etal_1988b, Hoffmann_etal_1989, Andresen_G_1994, Salamon_N_1998, Salamon_etal_2001, Schaller_etal_2001} and also by Beretta \cite{Beretta_1986, Beretta_1987, Beretta_2008}, Di\'osi and co-workers \cite{Diosi_etal_1996}, Crooks and Feng \cite{Crooks_2007, Feng_C_2009} and Brody and Hook \cite{Brody_H_2009}.  Some theoretical aspects of the adopted Riemannian formulation are discussed in Appendix \ref{Apx1}.  It must be noted that the Riemannian formulation replaces -- it cannot be used in conjunction with -- the traditional convex or concave hypersurface interpretation normally used in thermodynamics and information theory \cite{Andresen_etal_1988a}.

Firstly, the Riemannian geometric interpretation provides an {\it intrinsic differential}\index{intrinsic differential} or {\it line element}\index{line element} (its square, a {\it metric})\index{Riemannian metric} with which to measure distances along a specified path on the manifold \cite{Kreysig_1991, Zwillinger_2003}\footnote{Strictly, this line element is not a first fundamental form in Riemannian geometry \cite{Kreysig_1991, Zwillinger_2003}; its use as a distance measure is discussed in Appendix \ref{Apx1}.}:
\begin{align}
ds_{\mathfrak{H}^*} &= \sqrt{ {d ^2 \mathfrak{H}^*}}
= \sqrt{\sum\limits_{m,r=1}^R {d \langle {f_m } \rangle} \, g_{mr} \, {d \langle {f_r } \rangle}} 
= \sqrt{ d \vect{f}^{\top} \, \G \, d \vect{f}}  , 
\label{eq:ds_Hstar} \\
ds_{\phi} &= \sqrt{ {d ^2 \phi}}
= \sqrt {\sum\limits_{m,r=1}^R  {d \lambda_m} \, \gamma_{mr} \, {d \lambda_r}} 
= \sqrt{ d \vect{\Lambda}^{\top} \, \GAMMA \, d \vect{\Lambda}}.
\label{eq:ds_phi} 
\end{align}
where $\vect{f}=[\langle f_1 \rangle, \langle f_2 \rangle,...,\langle f_s \rangle]^{\top}$ and $\vect{\Lambda}=[\lambda_1,\lambda_2,...,\lambda_s]^{\top}$. Integration between points $a$ and $b$ along a path on the manifold, defined by the set of increments $d \vect{f}$ or $d \vect{\Lambda}$, gives the {\it arc length}\index{arc length} along that path between those points \cite{Kreysig_1991, Zwillinger_2003, Salamon_B_1983, Crooks_2007}:
\begin{align}
L_{\mathfrak{H}^*} &= \int\limits_{a}^b ds_{\mathfrak{H}^*}  
= \int\limits_{a}^b \sqrt{\sum\limits_{m,r=1}^R {d \langle {f_m } \rangle} \, g_{mr} \, {d \langle {f_r } \rangle}} 
= \int\limits_{a}^b \sqrt{ d \vect{f}^{\top} \, \G \, d \vect{f}}  , 
\label{eq:L1_Hstar} \\
L_{\phi} &= \int\limits_{a}^b ds_{\phi} 
= \int\limits_{a}^b \sqrt {\sum\limits_{m,r=1}^R  {d \lambda_m} \, \gamma_{mr} \, {d \lambda_r}} 
= \int\limits_{a}^b \sqrt{ d \vect{\Lambda}^{\top} \, \GAMMA \, d \vect{\Lambda}}.
\label{eq:L1_phi} 
\end{align}
The shortest such path is known as the {\it geodesic}\index{geodesic}. %, is shown in Figure \ref{fig:geom} for the surface $\mathfrak{H}^*(\langle f_m \rangle, \langle f_r \rangle)$. 
An infinite number of other paths on the manifold are also possible, of longer arc length, as also given by \eqref{eq:L1_Hstar} or \eqref{eq:L1_phi}.  If the manifold is parameterised by some parameter $\xi$ -- which can, but need not, correspond to time $t$ -- the arc lengths can be expressed in continuous form as:
\begin{align}
L_{\mathfrak{H}^*} &=\int\limits_0^{\xi_{max}} \sqrt{ {\sum\limits_{m,r=1}^R \frac{d \langle {f_m } \rangle}{d\xi} g_{mr}  \frac{d \langle {f_r } \rangle}{d\xi}}} \, d\xi 
= \int\limits_0^{\xi_{max}} \sqrt{ \vect{\dot{f}}^{\top} \, \G \, \vect{\dot{f}}} \, d\xi , 
\label{eq:L2_Hstar} \\
L_{\phi} &= \int\limits_0^{\xi_{max}} \sqrt{  {\sum\limits_{m,r=1}^R  \frac{d \lambda_m}{d\xi} \gamma_{mr}  \frac{d \lambda_r}{d\xi}}} \, d\xi 
= \int\limits_0^{\xi_{max}} \sqrt{ \vect{\dot{\Lambda}}^{\top} \, \GAMMA \, \vect{\dot{\Lambda}}} \, d\xi 
\label{eq:L2_phi} 
\end{align}
where the overdot indicates differentiation with respect to $\xi$. 

The symmetry of the Legendre transformation\index{Legendre transformation} \eqref{eq:Legendre} also permits a further insight. From \eqref{eq:MaxREnt_deriv2} and \eqref{eq:Massieu_deriv2}, the metrics $g_{mr}$ or $\gamma_{mr}$ within the intrinsic differentials \eqref{eq:ds_Hstar}-\eqref{eq:ds_phi} can be substituted respectively by $-\partial \lambda_r / \partial \langle f_m \rangle$ or $-\partial \langle f_r \rangle / \partial \lambda_m$, to give:
\begin{align}
ds_{\mathfrak{H}^*} 
&= \sqrt{-\sum\limits_{m,r=1}^R {d \langle {f_m } \rangle} \, \frac{\partial \lambda_r} {\partial \langle f_m \rangle} \, {d \langle {f_r } \rangle}} 
= \sqrt{-\sum\limits_{r=1}^R  d \lambda_r \, {d \langle {f_r } \rangle}} 
= \sqrt{- d \vect{\Lambda} \cdot d \vect{f}}
\label{eq:ds_Hstar2} \\
ds_{\phi} 
&=  \sqrt {-\sum\limits_{m,r=1}^R  {d \lambda_m} \, \frac{\partial \langle f_r \rangle}{\partial \lambda_m} \, {d \lambda_r}} 
=  \sqrt {-\sum\limits_{r=1}^R  {d \lambda_r} \, {d \langle f_r \rangle}} 
= \sqrt{- d \vect{\Lambda} \cdot d \vect{f}}
\label{eq:ds_phi2} 
\end{align}
In consequence, the intrinsic differentials are equal, $ds=ds_{\mathfrak{H}^*} = ds_{\phi}$, and so too are the arc lengths: 
\begin{equation}
L=L_{\mathfrak{H}^*} = L_{\phi} = \int\limits_0^{\xi_{max}} \sqrt{- \vect{\dot{\Lambda}} \cdot \vect{\dot{f}}} \; d\xi , 
\label{eq:L_equal}
\end{equation}
From a Riemannian geometric perspective, it therefore does not matter whether one examines a system using its $\mathfrak{H}^*(\langle f_1 \rangle, \langle f_2 \rangle, ...)$ or $\phi(\lambda_1, \lambda_2, ...)$ representation. The above identities -- touched on by several workers \cite{Levine_1986, Diosi_etal_1996, Salamon_N_1998, Crooks_2007} -- are not surprising, since the Legendre transforms $\mathfrak{H}^*$ and $\phi$ both have the character of entropy-related quantities, respectively indicating the (generic) entropy of a system and the capacity of a system to generate (generic) entropy \cite{Niven_MEP}. The quantity $- d \vect{\Lambda} \cdot d \vect{f}$ therefore expresses the second differential of generic entropy produced due to incremental changes in $\vect{\Lambda}$ and $\vect{f}$ (a generalised force-displacement relation). For all changes, $- d \vect{\Lambda} \cdot d \vect{f} \ge 0$ must be valid, to preserve a positive definite metric (whence $\protect{-{\vect{\dot{\Lambda}} \cdot \vect{\dot{f}}} \ge 0}$) %, hence an increase in $\vect{\dot{\Lambda}}$ causes a decrease in $\vect{\dot{f}}$, or vice versa) 
\cite{Weinhold_1975c}. This is in sympathy with a generalised form of the second law of thermodynamics, namely ``{\it each net mean increment of (generic) entropy produced along a path must be positive}''.\index{second law} 
%For a non-spontaneous (forced) change, for which $d \vect{\Lambda} \cdot d \vect{f} < 0$ (whence ${\vect{\dot{\Lambda}} \cdot \vect{\dot{f}}} < 0$), it is necessary to insert the factor $e=\text{sign}(\vect{\dot{\Lambda}} \cdot \vect{\dot{f}})= -1$ beneath the square root, so that the arc length $L = \int\nolimits_0^{\xi_{max}} \sqrt{e \; \vect{\dot{\Lambda}} \cdot \vect{\dot{f}}} \; d\xi$ remains positive \cite{Kreysig_1991, Zwillinger_2003}.
%It also imposes certain stability conditions between diagonal terms within the metric matrix \cite{Weinhold_1975c}.

One further consideration arises from the recognition that most probabilistic systems involve {\it quantised} phenomena, which can only be approximated by the above continuous representation.  For a system capable only of discrete increments in line elements $\Delta s_{\mathfrak{H}^*}$ or $\Delta s_{\phi}$ associated with a minimum dissipation parameter\index{minimum dissipation parameter} $\Delta \xi$ (e.g. a minimum dissipation time if $\xi=t$), the arc lengths\index{arc length} are more appropriately given as \cite{Nulton_etal_1985}:
\begin{align}
L_{\mathfrak{H}^*} &= \sum\limits_{\stepindex=1}^M \Delta s_{\mathfrak{H}^*,\stepindex}  
= \sum\limits_{\stepindex=1}^M \sqrt{ \Delta {\vect{f}_{\stepindex}}^{\top} \, \G_{\stepindex} \, \Delta \vect{f}_{\stepindex}}  
= \sum\limits_{\stepindex=1}^M \sqrt{ {\vect{\dot{f}}_{\stepindex}}^{\top} \, \G_{\stepindex} \, \vect{\dot{f}}_{\stepindex}} \; \Delta \xi_{\stepindex} , 
\label{eq:L_Hstar_disc} \\
L_{\phi} &=\sum\limits_{\stepindex=1}^M \Delta s_{\phi,\stepindex} 
= \sum\limits_{\stepindex=1}^M \sqrt{ \Delta {\vect{\Lambda}_{\stepindex}}^{\top} \, \GAMMA_{\stepindex} \, \Delta \vect{\Lambda}_{\stepindex}}
= \sum\limits_{\stepindex=1}^M \sqrt{ {\vect{\dot{\Lambda}}_{\stepindex}}^{\top} \, \GAMMA_{\stepindex} \, \vect{\dot{\Lambda}}_{\stepindex}} \; \Delta \xi_{\stepindex}
\label{eq:L_phi_disc} 
\end{align}
where $\stepindex$ is the index of each increment. The last terms in \eqref{eq:L_Hstar_disc}-\eqref{eq:L_phi_disc} invoke the finite difference forms $\vect{\dot{f}}_{\stepindex} = \Delta {\vect{f}_{\stepindex}}/\Delta \xi_{\stepindex}$ or $\vect{\dot{\Lambda}}_{\stepindex} = \Delta {\vect{\Lambda}_{\stepindex}}/\Delta \xi_{\stepindex}$, strictly valid only in the limits $\Delta \xi_{\stepindex} \to 0$. The two discrete length scales \eqref{eq:L_Hstar_disc}-\eqref{eq:L_phi_disc} are again equivalent, but there will most likely be some discrepancy between their values due to their finite difference formulation.

\subsection{\label{Action}Generalised Action Concepts and Least Action Bound} 
\begin{figure}[h]
%\begin{center}
\setlength{\unitlength}{0.6pt}
  \begin{picture}(500,350)
   \put(110,0){\includegraphics[width=65mm]{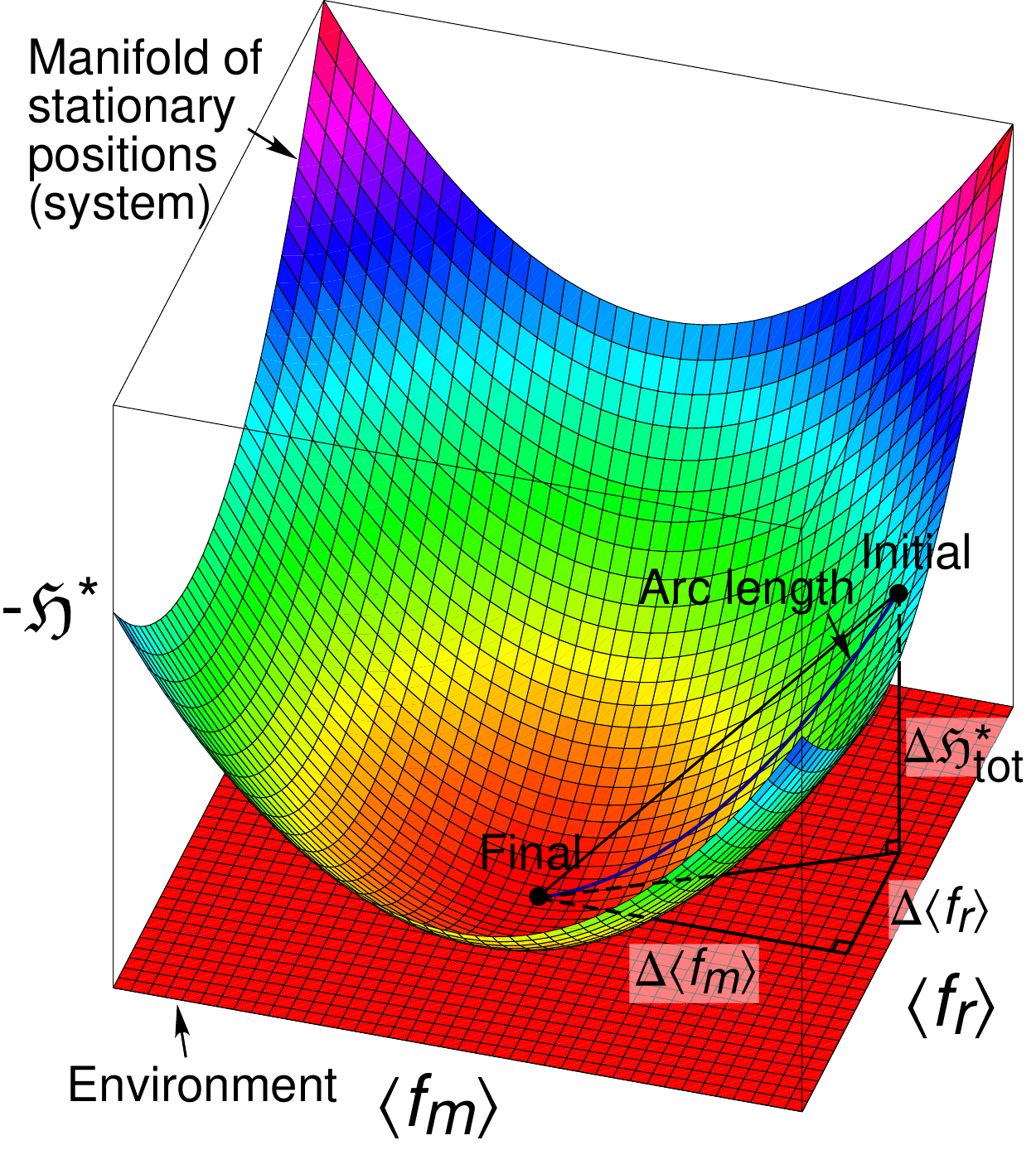}}
  \end{picture}
%\end{center}
\caption{Illustration of Riemannian geometry concepts, for a two-constraint system represented by $\mathfrak{H}^*(\langle f_m \rangle, \langle f_r \rangle)$ (for convenience the environment is shown as horizontal).}
\label{fig:geom}
\end{figure}

A Riemannian geometry can also be examined from a different perspective \cite{Salamon_A_G_B_1980, Salamon_B_1983, Nulton_etal_1985, Andresen_G_1994, Crooks_2007}, discussed with reference to Figure \ref{fig:geom}; the following analysis largely follows \cite{Nulton_etal_1985}, converted into generic form. Although applied to $\mathfrak{H}^*$, an analogous derivation can be given for the $\phi$ representation.  Consider a system on the manifold of stationary positions\index{manifold!of stationary positions}, subject to displacements $\{\Delta \langle f_r \rangle \}$ in its stationary position. The modified (generic) entropy $\mathfrak{H}^*(\{ \langle f_r \rangle + \Delta \langle f_r \rangle \})$ of the system can be expanded in a Taylor series about $\mathfrak{H}^*(\{ \langle f_r \rangle \})$, whence:
\begin{align}
\begin{split}
- \Delta \mathfrak{H}^*_{sys} &= \mathfrak{H}^*(\{ \langle f_r \rangle + \Delta \langle f_r \rangle \}) - \mathfrak{H}^*(\{ \langle f_r \rangle \}) 
%\\&
= \sum\limits_{r=1}^R \lambda_r \rvert_{\{\langle f_r \rangle\}} \Delta \langle f_r \rangle 
 + \frac{1}{2!}  \sum\limits_{m,r=1}^R \frac{\partial^2 \mathfrak{H}^*}{\partial \langle f_m \rangle \partial \langle f_r \rangle} \Biggr\rvert_{\{\langle f_r \rangle\}} {\Delta \langle f_m \rangle} {\Delta \langle f_r \rangle}  
\\ &
+ \frac{1}{3!}  \sum\limits_{m,r,\ell=1}^R \frac{\partial^3 \mathfrak{H}^*}{\partial \langle f_{\ell} \rangle \partial \langle f_m \rangle \partial \langle f_r \rangle} \Biggr\rvert_{\{\langle f_r \rangle\}} {\Delta \langle f_{\ell} \rangle} {\Delta \langle f_m \rangle} {\Delta \langle f_r \rangle} + ...
\label{eq:H_Taylor1}
\end{split}
\end{align}
where $\Delta \mathfrak{H}^*_{sys}$ is the net {\it increase} in entropy \cite{Nulton_etal_1985}, in which use is made of \eqref{eq:MaxREnt_deriv1}. The corresponding change in entropy of the ``reservoir'' or ``environment'' of constant $\{ \lambda_r^{env} \}$, by which this change is effected, is given (exactly) by \cite{Salamon_B_1983, Nulton_etal_1985}:
\begin{align}
- \Delta \mathfrak{H}^{env} =
\mathfrak{H}^{env}(\{ \langle f_r \rangle + \Delta \langle f_r \rangle \}) 
- \mathfrak{H}^{env}(\{ \langle f_r \rangle \})
= \sum\limits_{r=1}^R \lambda_r^{env} \rvert_{\{\langle f_r \rangle\}} \Delta \langle f_r \rangle^{env} 
\label{eq:tangent}
\end{align}
At the stationary state, $\lambda_r = \lambda_r^{env}$, whilst from the constraints (conservation laws), $\Delta \langle f_r \rangle=-\Delta \langle f_r \rangle^{env}$ \cite{Nulton_etal_1985}.  Addition of \eqref{eq:H_Taylor1} and \eqref{eq:tangent} thus yields the total change in the entropy of the system and environment for the step process: 
\begin{align}
\begin{split}
-\Delta \mathfrak{H}^* &= \frac{1}{2!}  \sum\limits_{m,r=1}^R \frac{\partial^2 \mathfrak{H}^*}{\partial \langle f_m \rangle \partial \langle f_r \rangle} \Biggr\rvert_{\{\langle f_r \rangle\}} {\Delta \langle f_m \rangle} {\Delta \langle f_r \rangle}  
%\\ & 
+\frac{1}{3!}  \sum\limits_{m,r,\ell=1}^R \frac{\partial^3 \mathfrak{H}^*}{\partial \langle f_{\ell} \rangle \partial \langle f_m \rangle \partial \langle f_r \rangle} \Biggr\rvert_{\{\langle f_r \rangle\}} {\Delta \langle f_{\ell} \rangle} {\Delta \langle f_m \rangle} {\Delta \langle f_r \rangle} + ...
\end{split}
\label{eq:H_Taylor_diff}
\end{align}
Provided the manifold is smooth, continuous, continuously differentiable (i.e., there are no phase changes in the neighbourhood) and the step sizes $\{ \Delta \langle f_r \rangle \}$ are small, we can neglect the higher order terms in \eqref{eq:H_Taylor_diff}, giving:
\begin{align}
\Delta \mathfrak{H}^*_{\stepindex} &\approx \frac{1}{2}  \sum\limits_{m,r=1}^R {\Delta \langle f_m \rangle_{\stepindex}} \; g_{mr,\stepindex} \rvert_{\{\langle f_r \rangle\}} \; {\Delta \langle f_r \rangle_{\stepindex}}  
= \frac{1}{2} \Delta {\vect{f}_{\stepindex}}^{\top} \, \G_{\stepindex} \, \Delta \vect{f}_{\stepindex}
\label{eq:H_Taylor_diff2}
\end{align}
where the subscript denotes the $\stepindex$th equilibration step. In the $\phi$ representation, the analogous form is obtained:
\begin{align}
- \Delta \phi_{\stepindex} &\approx \frac{1}{2}  \sum\limits_{m,r=1}^R {\Delta \lambda_{m,{\stepindex}}} \; \gamma_{mr,{\stepindex}} \rvert_{\{\lambda_r\}} \; {\Delta \lambda_{r,{\stepindex}}}  
= \frac{1}{2} \Delta {\vect{\Lambda}_{\stepindex}}^{\top} \, \GAMMA_{\stepindex} \, \Delta \vect{\Lambda}_{\stepindex} 
\label{eq:phi_Taylor_diff2}
\end{align}
The total increase in entropy or decrease in potential of the system and environment subject to a $M$-step process is therefore:
\begin{align}
\Delta \mathfrak{H}^*_{tot} &=\sum\limits_{\stepindex=1}^M \Delta \mathfrak{H}^*_{\stepindex} 
\approx \sum\limits_{\stepindex=1}^M \frac{1}{2} \Delta {\vect{f}_{\stepindex}}^{\top} \, \G_{\stepindex} \, \Delta \vect{f}_{\stepindex}  
= \sum\limits_{\stepindex=1}^M \frac{1}{2} { {\vect{\dot{f}}_{\stepindex}}^{\top} \, \G_{\stepindex} \, \vect{\dot{f}}_{\stepindex}} \; \Delta \xi_{\stepindex} \Delta \xi_{\stepindex} , 
\label{eq:deltaH_tot}
\\
-\Delta \phi_{tot} &= - \sum\limits_{\stepindex=1}^M \Delta \phi_{\stepindex} 
\approx \sum\limits_{\stepindex=1}^M \frac{1}{2} \Delta {\vect{\Lambda}_{\stepindex}}^{\top} \, \GAMMA_{\stepindex} \, \Delta \vect{\Lambda}_{\stepindex}
= \sum\limits_{\stepindex=1}^M \frac{1}{2} { {\vect{\dot{\Lambda}}_{\stepindex}}^{\top} \, \GAMMA_{\stepindex} \, \vect{\dot{\Lambda}}_{\stepindex}} \; \Delta \xi_{\stepindex} \Delta \xi_{\stepindex}
\label{eq:deltaphi_tot}
\end{align}
Recognising $\Delta \xi_{\stepindex}$ as the minimum dissipation parameter for the $\stepindex$th step (e.g.\ the minimum dissipation time if $\xi=t$), one such term may be factored out of each sum, to give the {\it mean minimum dissipation parameter}\index{mean minimum dissipation parameter} $\bar{\epsilon}_{n}$ for $n \in \{\mathfrak{H}^*, \phi\}$ \cite{Salamon_B_1983, Nulton_etal_1985}.  This gives $\Delta \mathfrak{H}^*_{tot} = \bar{\epsilon}_{\mathfrak{H}^*}  \mathcal{J}_{\mathfrak{H}^*}$ or $-\Delta \phi_{tot}= \bar{\epsilon}_{\phi} \mathcal{J}_{\phi}$, with:
\begin{align}
\mathcal{J}_{\mathfrak{H}^*} 
= \sum\limits_{\stepindex=1}^M \frac{1}{2} {{\vect{\dot{f}}_{\stepindex}}^{\top} \, \G_{\stepindex} \, \vect{\dot{f}}_{\stepindex}} \; \Delta \xi_{\stepindex} , 
\label{eq:J_H}
\\
\mathcal{J}_{\phi} 
= \sum\limits_{\stepindex=1}^M \frac{1}{2} {{\vect{\dot{\Lambda}}_{\stepindex}}^{\top} \, \GAMMA_{\stepindex} \, \vect{\dot{\Lambda}}_{\stepindex}} \; \Delta \xi_{\stepindex}
\label{eq:J_phi}
\end{align}
The summands $\frac{1}{2} {{\vect{\dot{f}}_{\stepindex}}^{\top} \, \G_{\stepindex} \, \vect{\dot{f}}_{\stepindex}}$ or $\frac{1}{2} {{\vect{\dot{\Lambda}}_{\stepindex}}^{\top} \, \GAMMA_{\stepindex} \, \vect{\dot{\Lambda}}_{\stepindex}}$ in \eqref{eq:J_H}-\eqref{eq:J_phi} can be viewed as {\it generalised energy} terms, akin to the kinetic energy in mechanics, with the metric $\G_{\stepindex}$ or $\GAMMA_{\stepindex}$ representing the ``mass'' and $\vect{\dot{f}}_{\stepindex}$ or $\vect{\dot{\Lambda}}_{\stepindex}$ the ``velocity'' \cite{Lanczos_1966}. The terms $\mathcal{J}_{\mathfrak{H}^*}$ or $\mathcal{J}_{\phi}$ can then be interpreted as the discrete {\it generalised action}\index{generalised action}\index{action} of the specified process \cite{Crooks_2007}, again by analogy with mechanics\footnote{Crooks \cite{Crooks_2007} applies the terms ``energy" and ``action" interchangeably; we consider that the present definitions are more in keeping with those used in mechanics. Many authors include the $\bar{\epsilon}_n$ term within $\mathcal{J}_n$, but we here wish to preserve the mathematical structure of a generalised action principle.}. From the previous considerations (\S\ref{Riemann}), the two action sums are equivalent, although once again, discrepancies may emerge from their finite difference formulation.

From the discrete form of the Cauchy-Schwarz inequality\index{Cauchy-Schwarz inequality}:
\begin{equation}
\biggl(\sum\nolimits_{\stepindex=1}^M {a_\stepindex}^2 \biggr) \biggl(\sum\nolimits_{\stepindex=1}^M {b_\stepindex}^2 \biggr) \ge \biggl(\sum\nolimits_{\stepindex=1}^M a_\stepindex b_\stepindex \biggr)^2
\label{eq:CS}
\end{equation}
with $a_\stepindex=\sqrt{ \vect{\dot{f}}^{\top} \, \G_{\stepindex} \, \vect{\dot{f}}} \; \Delta \xi_{\stepindex}$ or $\sqrt{\vect{\dot{\Lambda}}^{\top} \, \GAMMA_{\stepindex} \, \vect{\dot{\Lambda}}} \; \Delta \xi_{\stepindex}$ and $b_\stepindex=1$, it can be shown that \cite{Nulton_etal_1985}:
\begin{equation}
\bar{\epsilon}_n \mathcal{J}_n \ge \frac {L_n^2}{2M}
\label{eq:leastaction1}
\end{equation}
Physically, the number of steps is equal to $M=\xi_{max}/\bar{\epsilon}_{n}$, whence \eqref{eq:leastaction1} reduces to \cite{Salamon_B_1983, Nulton_etal_1985}:
\begin{equation}
\bar{\epsilon}_n \mathcal{J}_n \ge \frac {\bar{\epsilon}_n L_n^2}{2 \, \xi_{max}}
\qquad \text{or} \qquad
\mathcal{J}_n \ge \frac {L_n^2}{2 \, \xi_{max}}
\label{eq:leastaction2}
\end{equation}
Eqs.\ \eqref{eq:leastaction1}-\eqref{eq:leastaction2} can be considered a generalised {\it least action bound}\index{least action bound}\index{principle of least action} \cite{Crooks_2007}, applicable to all probabilistic systems amenable to analysis by Jaynes' method. Its physical interpretation is that it specifies the minimum cost or penalty\index{minimum cost, entropy units}\index{lower bound, entropy cost}, in units of dimensionless entropy per unit $\xi$, to move the system from one stationary position (at $\xi=0$) to another (at $\xi=\xi_{max}$) along the given path at the specified rates $\vect{\dot{\Lambda}}$ and/or $\vect{\dot{f}}$.  If the latter rates proceed infinitely slowly, the lower bound of the action is zero, indicating that the process can be conducted at zero cost; otherwise, it is necessary to ``do generalised work'' to move the system along the manifold of stationary positions within a finite parameter duration $\xi_{max}$. 

The generalised least action bound thus provides a lower bound for the ``transition cost'' of a process (in entropy-related units). If the process is reversible, the cost would be zero, but no process can be reversible in practice. Identification of this minimum cost is of paramount importance: there is no point in undertaking expensive changes to the process, or initiating costly social or political changes, in the attempt to do better than the minimum predicted by \eqref{eq:leastaction1}-\eqref{eq:leastaction2}. Taking a thermodynamic example, the method can be applied to determine the minimum cost of industrial processes such as work extraction from combustion, a question of fundamental importance to human society.  Most thermodynamics and engineering textbooks give the Carnot limit as the theoretical limit of efficiency, but the limits imposed by finite time thermodynamics are more restrictive (see \S\ref{Equil}).

The generalised least action bound therefore emerges from the Riemannian geometry of the state space, and hence from somewhat different considerations than the principle of least action employed in mechanics \cite{Lanczos_1966}.  We consider that the two principles are connected, but are unable to examine this topic further here.  For further exploratory expositions, the reader is referred to the work of Crooks \cite{Crooks_2007}, Caticha \cite{Caticha_Cafaro_2007} and Wang \cite{Wang_2005, Wang_2006}. 

The above discrete sums \eqref{eq:deltaH_tot}-\eqref{eq:deltaphi_tot} can also be presented in integral form. Consider a system represented by $\mathfrak{H}^*$, subjected to a finite change in the multipliers $\Delta \lambda_r$ due to movement of the reference environment. The incremental change in entropy is, again to first order (compare \eqref{eq:H_Taylor_diff} and discussion after \eqref{eq:L_equal}) \cite{Salamon_B_1983, Nulton_etal_1985, Andresen_G_1994, Crooks_2007}:
\begin{align}
d \mathfrak{H}^* 
&\approx - \frac{1}{2!}  \sum\limits_{r=1}^R {\Delta \lambda_r} {d \langle f_r \rangle}  
\label{eq:H_Taylor_diff_cts}
\end{align}
Substituting $\Delta \lambda_r = - \sum\limits_{m=1}^R g_{mr} \Delta \langle f_m \rangle$ from \eqref{eq:MaxREnt_deriv2}, and assuming a first order decay process:
\begin{equation}
\langle \dot{f}_m \rangle = \frac{\langle {f}_m \rangle - \langle {f}_m \rangle_{env}}{\epsilon_{\mathfrak{H}^*}} = \frac{\Delta \langle {f}_m \rangle}{\epsilon_{\mathfrak{H}^*}} 
\label{eq:dissip_param}
\end{equation}
where $\epsilon_{\mathfrak{H}^*}$ is a minimum dissipation parameter (reciprocal rate constant), \eqref{eq:H_Taylor_diff_cts} yields:
\begin{align}
d \mathfrak{H}^* 
&=  \frac{1}{2} \sum\limits_{m,r=1}^R {\langle \dot{f}_m \rangle} \, g_{mr} \, {d \langle f_r \rangle} \; \epsilon_{\mathfrak{H}^*}
\label{eq:H_Taylor_diff_cts2}
\end{align}
The total change in entropy $\Delta \mathfrak{H}^*_{tot} =  \int\nolimits_0^{\xi_{max}} d \mathfrak{H}^*$ is then obtained as\index{least action bound}:
\begin{align}
\Delta \mathfrak{H}^*_{tot} &
=  \int\limits_0^{\xi_{max}} \frac{1}{2} \; { \vect{\dot{f}}^{\top} \, \G \, \vect{\dot{f}}} \,  {\epsilon}_{\mathfrak{H}^*} \; d\xi 
= \bar{\epsilon}_{\mathfrak{H}^*}  \int\limits_0^{\xi_{max}} \frac{1}{2} \; { \vect{\dot{f}}^{\top} \, \G \, \vect{\dot{f}}} \, d\xi 
= \bar{\epsilon}_{\mathfrak{H}^*} \mathcal{J}_{\mathfrak{H}^*}
\label{eq:deltaH_tot_int}
\end{align}
Similarly, in the $\phi$ representation, we obtain:
\begin{align}
-\Delta \phi_{tot} &
= \int\limits_0^{\xi_{max}} \frac{1}{2} \; { \vect{\dot{\Lambda}}^{\top} \, \GAMMA \, \vect{\dot{\Lambda}}} \, {\epsilon}_{\phi} \;  d\xi 
= \bar{\epsilon}_{\phi} \int\limits_0^{\xi_{max}} \frac{1}{2} \; { \vect{\dot{\Lambda}}^{\top} \, \GAMMA \, \vect{\dot{\Lambda}}} \, d\xi 
= \bar{\epsilon}_{\phi} \mathcal{J}_{\phi}
\label{eq:deltaphi_tot_int}
\end{align}
In the continuous representation, the process does not proceed by a series of finite steps; instead, the reference variables continuously move ahead of those of the system \cite{Salamon_B_1983, Nulton_etal_1985, Andresen_G_1994}.  However, we still see the influence of a finite decay parameter $\epsilon_n$, which on integration yields the mean parameter $\bar{\epsilon}_n$. Each $\mathcal{J}_n$ term above can be regarded as the {\it action integral} corresponding respectively to \eqref{eq:J_H}-\eqref{eq:J_phi}.  Based on the integral form of the Cauchy-Schwarz inequality \cite{Zwillinger_2003}, it can be shown that the integral actions also satisfy the least action bound\index{least action bound} \eqref{eq:leastaction1}-\eqref{eq:leastaction2}, with $L_n$ in integral form \cite{Salamon_B_1983, Nulton_etal_1985, Andresen_G_1994, Crooks_2007}. 

Finally, for the least action bound \eqref{eq:leastaction2} to achieve equality, the summands or integrands of the arc length $L_n$ and action $J_n$ must be constant. This gives the simple result that for slow processes with constant dissipation parameter $\epsilon_n=\bar{\epsilon}_n$, the minimum action (whence minimum in $\bar{\epsilon}_n J_n$) is attained by a process which proceeds at a constant speed $\dot{s}_n= \sqrt{ \vect{\dot{f}}^{\top} \, \G \, \vect{\dot{f}}} = \sqrt{ \vect{\dot{\Lambda}}^{\top} \, \GAMMA \, \vect{\dot{\Lambda}}}$ \cite{Nulton_S_1988, Salamon_etal_1988, Andresen_G_1994}. For a constant metric, this is equivalent to constant rates of change of the parameter vector $\vect{\dot{f}}$ and/or $\vect{\dot{\Lambda}}$. For systems with a variable dissipation parameter $\epsilon(\xi)$, it was first considered that the minimum is attained at the constant speed $ds/d\eta$, expressed in the ``natural'' parameter units $\eta = \xi/\epsilon$ \cite{Nulton_S_1988, Salamon_etal_1988, Andresen_G_1994, Diosi_etal_1996}. This however oversimplifies the minimisation problem, which is better handled within a discrete (stepwise) framework \cite{Nulton_etal_1985, Salamon_etal_2001}. As discussed in \S\ref{Equil}, such principles have been widely applied to thermodynamic systems. 

\subsection{\label{Geodesic}Minimum Path Length Principle} 
The above discrete or continuous forms of the least action bound \eqref{eq:leastaction1}-\eqref{eq:leastaction2} are based on consideration of a specified path on the manifold of stationary positions, of arc length $L_n$.  In many situations, we may wish to determine the path of minimum arc length $L_{n,min}$ -- the {\it geodesic}\index{geodesic}\index{Riemannian geodesic} -- on the manifold of stationary positions. From the calculus of variations, this is given by the Euler-Lagrange equations\index{Euler-Lagrange equations} \cite{Weinstock_1974}: 
\begin{align}
\frac{\partial \dot{s}_{\mathfrak{H}^*}}{\partial \vect{f}} - \frac{d}{d \xi} \frac{\partial \dot{s}_{\mathfrak{H}^*}}{\partial \vect{\dot{f}}} &= \vect{0} 
\label{eq:Euler_Lagr1}
\\
%&\hspace{-80pt}\text{or}&
\frac{\partial \dot{s}_{\phi}}{\partial \vect{\Lambda}} - \frac{d}{d \xi} \frac{\partial \dot{s}_{\phi}}{\partial \vect{\dot{\Lambda}}} &= \vect{0}
\label{eq:Euler_Lagr2}
\end{align}
where $\dot{s}_{\mathfrak{H}^*} = \sqrt{ \vect{\dot{f}}^{\top} \, \G \, \vect{\dot{f}}}$ and $\dot{s}_{\phi} = \sqrt{ \vect{\dot{\Lambda}}^{\top} \, \GAMMA \, \vect{\dot{\Lambda}}}$ 
are the integrands respectively of $L_{\mathfrak{H}^*}$ or $L_{\phi}$ \eqref{eq:L1_Hstar}-\eqref{eq:L2_phi}. % (whence from \eqref{eq:L_equal}, $\dot{s} = \sqrt{ \vect{\dot{\Lambda}} \cdot \vect{\dot{f}}}$). 
For two-dimensional parameters $\vect{f}, \vect{\Lambda} \in \mathbb{R}^2$, \eqref{eq:Euler_Lagr1}-\eqref{eq:Euler_Lagr2} can be reduced further in terms of the three unit normals to the surface, giving the curve(s) on the manifold for which the geodesic curvature vanishes \cite{Kreysig_1991, Zwillinger_2003, Brody_H_2009}. Depending on the specified problem, a geodesic may not exist, or there may be multiple or il-defined solutions. Provided it does exist, a geodesic leads to the double minimisation principle:
\begin{equation}
\mathcal{J}_n \ge \frac {L_n^2}{2 \, \xi_{max}} \ge \frac {{L_{n,min}}^2}{2 \, \xi_{max}}
\label{eq:doubleleastaction}
\end{equation}
where the right hand side indicates the absolute lower bound for the action, irrespective of path. This principle has been applied to thermodynamic systems, as will be discussed in \S\ref{Equil}.
%%############################################################################
\section{\label{Apps}Applications} 
%% ############################################################################
%%
As noted, the foregoing Riemannian geometric interpretation (\S\ref{Riemann}) has mainly been presented within an equilibrium thermodynamics context\index{finite time thermodynamics} \cite{Salamon_A_G_B_1980, Salamon_N_A_B_1980, Salamon_B_1983, Salamon_I_B_1983, Andresen_1983_book, Salamon_N_I_1984, Andresen_B_O_S_1984, Salamon_N_B_1985, Schlogl_1985, Feldman_etal_1985, Nulton_etal_1985, Feldman_etal_1986, Levine_1986, Nulton_S_1988, Salamon_etal_1988, Andresen_etal_1988a, Andresen_etal_1988b, Hoffmann_etal_1989, Andresen_G_1994, Salamon_N_1998, Salamon_etal_2001, Schaller_etal_2001, Beretta_1986, Beretta_1987, Beretta_2008, Diosi_etal_1996, Crooks_2007, Feng_C_2009, Brody_H_2009}, although it has been applied to non-equilibrium thermodynamic and flow systems \cite{Nathanson_S_1980, Gilmore_1982, Sien_B_1991, Chen_2001, Chen_2003}, information coding \cite{Flick_etal_1987} and in economics \cite{Salamon_K_A_N_1987}.  In the following sections, the utility of Riemannian geometric properties and the least action bound are demonstrated for two types of system: a thermodynamic system at equilibrium, and a flow system at steady state.     

\subsection{\label{Equil}Equilibrium Thermodynamic Systems}
The application of Riemannian geometric principles to equilibrium thermodynamic systems\index{equilibrium systems}\index{least action bound!equilibrium systems} has constituted a major new development over the past three decades, forming an important plank of {\it finite-parameter} or (with $\xi = t$) {\it finite-time thermodynamics}\index{finite time thermodynamics} \cite{Andresen_B_O_S_1984, Salamon_N_1998, Salamon_etal_2001}. Such analyses have progressed in four overlapping stages:
\begin{list}{$\bullet$}{\topsep 3pt \itemsep 3pt \parsep 0pt \leftmargin 8pt \rightmargin 0pt \listparindent 0pt
\itemindent 0pt}
\item The initial studies by Weinhold \cite{Weinhold_1975a, Weinhold_1975b, Weinhold_1975c, Weinhold_1975d, Weinhold_1976} and early work by Salamon, Andresen, Berry, Nulton and coworkers \cite{Salamon_A_G_B_1980, Salamon_B_1983, Salamon_I_B_1983, Salamon_N_I_1984, Nulton_etal_1985, Nulton_S_1988, Salamon_etal_1988} all examined a manifold based on an internal energy representation $U(X_1, X_2, ...)$, as a function of extensive variables $X_r$, which include the thermodynamic entropy $S$. The resulting quantity $\bar{\epsilon}_U \mathcal{J}_U$ (in the present notation) was interpreted as an {\it availability}\index{availability} or {\it exergy}\index{exergy} function, with \eqref{eq:leastaction2} indicating the most efficient path (defined by the minimum amount of work or minimum loss of availability) required to move the equilibrium position of the system \cite{Salamon_B_1983, Salamon_I_B_1983}. Such analyses complement the thermodynamic geometry used by Gibbs \cite{Gibbs_1873b, Gibbs_1875}, and fit well with the traditional heat-work framework of 19th century thermodynamics. 
%This approach has now been applied to the optimisation of many engineering and industrial processes, including engine cycles, heat engines and pumps, chemical reactors, distillation towers \cite{**}.
\item Subsequently, following earlier pioneering works \cite{Ruppeiner_1979, Nathanson_S_1980}, the entropy manifold $S(\tilde{X}_1, \tilde{X}_2, ...)$ was examined from a Riemannian perspective \cite{Salamon_N_I_1984, Andresen_B_O_S_1984, Schlogl_1985, Nulton_etal_1985, Andresen_G_1994, Andresen_etal_1988b, Hoffmann_etal_1989, Andresen_G_1994, Salamon_N_1998, Crooks_2007, Feng_C_2009}, where $\tilde{X}_r$ are the new extensive variables, of course related to the $U(X_1, X_2, ...)$ representation by Jacobian transformation \cite{Salamon_N_I_1984}.  The quantity $\bar{\epsilon}_S \mathcal{J}_S$ was interpreted as a measure of {\it energy dissipation}\index{energy dissipation} or {\it entropy production}\index{entropy production}, again providing a measure of process efficiency. It was realised that the lower bound in \eqref{eq:leastaction2} provides a formal, mathematical definition of the degree of irreversibility of a transition between equilibrium positions, with reversibility only for $\mathcal{J}_S = 0$ (a definition vastly preferable to the cumbersome word-play still used in thermodynamics references; see the scathing criticism by Truesdell \cite{Truesdell_1969}). However, the primacy of the entropy representation over that based on internal energy was not fully appreciated in these early studies. %, although it was noted that the $S$ form in many cases yields a diagonal metric $\G$. 
The applicability of Riemannian geometry in other contexts -- based directly on the MaxEnt framework of Jaynes \cite{Jaynes_1957, Jaynes_1963} -- is hinted at by Levine \cite{Levine_1986}, but unfortunately was not developed further at the time, nor, to the authors' knowledge, in any subsequent studies.  
\item Several studies have considered an entropy representation based on a metric defined on a probability space $\{p_i\}$, either from the Boltzmann principle \cite{Ruppeiner_1979} or using a Shannon or relative entropy measure \cite{Salamon_N_B_1985, Schlogl_1985, Feldman_etal_1985, Feldman_etal_1986, Levine_1986, Beretta_1986, Beretta_1987, Beretta_2008, Crooks_2007, Feng_C_2009, Brody_H_2009}. Several authors \cite{Beretta_1986, Beretta_1987, Beretta_2008, Crooks_2007, Feng_C_2009, Brody_H_2009}) have extended this analysis, to establish a connection with the Fisher information matrix \cite{Fisher_1922} and an ``entropy differential metric'' of Rao \cite{Rao_1945}. The analysis is also intimately connected with paths in a space of square root probabilities, and thence to formulations of quantum mechanics \cite{Beretta_1986, Beretta_1987, Beretta_2008}. These insights -- not examined further here -- demand further detailed attention; they may well furnish an explanation for the utility of extremisation methods based on the Fisher information function in many physical problems \cite{Frieden_2004}. 
\item Finally, several workers realised that Riemannian geometric principles can be applied to Legendre-transformed representations, e.g.\ based on various forms of the free energy $F$ (conjugate to $U$) or the negative Planck potential $F/T$ (conjugate to $S$), as functions of the intensive variables (or functions thereof) \cite{Schlogl_1985, Levine_1986, Crooks_2007, Feng_C_2009}. This approach offers particular advantages for the analysis of real thermodynamic systems, in which the control parameters tend to be intensive\index{intensive variables} rather than extensive variables\index{extensive variables} (the canonical ensemble), and for which the intensive variables do not exhibit sharp transitions or singularities associated with phase changes, as is the case for extensive variables \cite{Crooks_2007}. Furthermore, the resulting metric is equivalent to the variance-covariance matrix of the constraints \eqref{eq:Massieu_deriv2}, and is therefore connected to fluctuation-dissipation processes within the system.
\end{list}

For completeness, we demonstrate -- for a microcanonical thermodynamic system -- how Riemannian geometric properties emerge as an inherent feature of Jaynes' MaxEnt formulation.  Consider an isolated thermodynamic system, containing molecules of possible energy levels $\epsilon_i$ and volume elements $V_j$, subject to constraints on the mean energy $\langle U \rangle$ and mean volume $\langle V \rangle$. We consider the joint probability $p_{ij}$ of a particle simultaneously occupying an energy level and volume element, giving the entropy function:
\begin{align}
\mathfrak{H}_{eq} &=   -  \sum\limits_{i}   \sum\limits_{j}   p_{ij} \ln p_{ij} ,
\label{eq:Heq}
\end{align}
where, without knowledge of any additional influences, we assume that each joint level $ij$ is equally probable (hence the priors $q_{ij}$ cancel out). Eq.\ \eqref{eq:Heq} is maximised subject to the constraints:
\begin{gather}
\sum\limits_{i}   \sum\limits_{j}   p_{ij} = 1,
\label{eq:C0eq} 
\\
\sum\limits_{i}   \sum\limits_{j}   p_{ij} \epsilon_i = \langle U \rangle,
\label{eq:CUeq}
\\
\sum\limits_{i}   \sum\limits_{j}   p_{ij} V_j = \langle V \rangle,
\label{eq:CVeq}
\end{gather}
to give the equilibrium position:
\begin{gather}
%\begin{split}
p_{ij}^* = \dfrac { e^ {- \lambda_{U} U_i  - \lambda_V V_j}} {\sum\limits_{i}   \sum\limits_{j}  e^{- \lambda_{U} U_i  - \lambda_V V_j}} 
= \dfrac{1}{Z} {e^{- \lambda_{U} U_i  - \lambda_V V_j }},
%\end{split}
\label{eq:pstar2eq} 
\end{gather}
where $Z$ is the partition function. From the existing body of thermodynamics, we can identify the Lagrangian multipliers as $\lambda_{U}=1/kT$ and $\lambda_V = P/kT$, where $k$ is the Boltzmann constant, $T$ is absolute temperature and $P$ is absolute pressure. Eq.\ \eqref{eq:pstar2eq} and Jaynes' relations \eqref{eq:Hstar}-\eqref{eq:Legendre} and \eqref{eq:Massieu} then reduce to\index{thermodynamic entropy}\index{manifold! of equilibrium positions}:
\begin{gather}
p_{ij}^* = \dfrac{1}{Z} {e^ {- U_i / kT  - P V_j / kT}},
\label{eq:pstar2eq2} 
\\
S^* = k \mathfrak{H}^*_{eq} = k \ln Z  + \frac{\langle U \rangle}{T} + \frac{ P \langle V \rangle}{T}
\label{eq:Sstar}
\\
k \, \vect{\Lambda}_{eq} = \biggl[ \frac{\partial S^*}{\partial \langle U \rangle }, \frac{\partial S^*}{\partial \langle V \rangle} \biggr]^{\top} = \biggl[ \frac{1}{T}, \frac{P}{T} \biggr]^{\top}  
\label{eq:Sstar_deriv1}
\\
- k \, \G_{eq}  
= \begin{bmatrix} \dfrac{\partial^2 S^*}{\partial \langle U \rangle^2}, & \dfrac{\partial^2 S^*}{\partial \langle U \rangle \partial \langle V \rangle} \\
\dfrac{\partial^2 S^*}{\partial \langle V \rangle \partial \langle U \rangle}, &\dfrac{\partial^2 S^*}{\partial \langle V \rangle^2} \end{bmatrix} 
= \begin{bmatrix} \dfrac{\partial}{\partial \langle U \rangle} \biggl( \dfrac{1}{T} \biggr), & \dfrac{\partial}{\partial \langle U \rangle} \biggl( \dfrac{P}{T} \biggr)\\
\dfrac{\partial}{\partial \langle V \rangle} \biggl( \dfrac{1}{T} \biggr), & \dfrac{\partial}{\partial \langle V \rangle} \biggl( \dfrac{P}{T} \biggr) \end{bmatrix}
\label{eq:Sstar_deriv2}
\\
\psi = k \phi_{eq}  =  -k \ln Z =- S^*  +  \frac{\langle U \rangle}{T} + \frac{ P \langle V \rangle}{T} = \frac{G}{T}
\label{eq:psi}
\\
\vect{f}_{eq} = 
\biggl[ \frac{\partial \psi}{\partial (\frac{1}{T})}, \frac{\partial \psi}{\partial (\frac{P}{T})} \biggr]^{\top} = \bigl[ \langle U \rangle, \langle V \rangle \bigr]^{\top}
\label{eq:psi_deriv1}
\\
- \frac{\GAMMA_{eq}}{k} 
= \begin{bmatrix} \dfrac{\partial^2 \psi}{\partial (\frac{1}{T})^2}, & \dfrac{\partial^2 \psi}{\partial (\frac{1}{T}) \partial (\frac{P}{T})} \\
\dfrac{\partial^2 \psi}{\partial (\frac{P}{T}) \partial (\frac{1}{T})}, &\dfrac{\partial^2 \psi}{\partial (\frac{P}{T})^2} \end{bmatrix}
= \begin{bmatrix} \dfrac{\partial \langle U \rangle}{\partial (\frac{1}{T})}, & \dfrac{\partial \langle V \rangle}{\partial (\frac{1}{T}) } \\
\dfrac{\partial \langle U \rangle}{\partial (\frac{P}{T}) }, &\dfrac{\partial \langle V \rangle}{\partial (\frac{P}{T})} \end{bmatrix} 
\label{eq:psi_deriv2}
\\
k \, \G_{eq} \; \frac{\GAMMA_{eq}}{k} = \I
\label{eq:Legendre_eq}
\end{gather}
where $S^*$ is the thermodynamic entropy at an equilibrium position, $\psi$ is the negative Planck potential \cite{Planck_1922, Planck_1932} (negative Massieu function \cite{Massieu_1869}) and $G$ is the Gibbs free energy. By Jacobian transformation of variables, using the following material properties (susceptibilities)\index{susceptibilities!equilibrium systems} \cite{Callen_1960, Callen_1985, Weinhold_1975c}:
\begin{alignat}{2}
\hspace{5pt} &\text{Heat capacity at constant pressure:}\hspace{15pt} &C_P &= \biggl( \frac{\partial \langle H \rangle}{\partial T} \biggr)_{P} 
\label{eq:C_P}\\
&\text{Isothermal compressibility:} &\compress_T &= - \frac {1}{\langle V \rangle} \biggl( \frac{\partial \langle V \rangle}{\partial P} \biggr)_{T}   
\label{eq:kappa_T}\\
&\text{Coefficient of thermal expansion:} &\alpha &= \frac {1}{\langle V \rangle} \biggl( \frac{\partial \langle V \rangle}{\partial T} \biggr)_{P}  
\label{eq:alpha}
\end{alignat}
where $\langle H \rangle=\langle U \rangle + P \langle V \rangle$ is the enthalpy, as well as the equality of cross-derivatives (Maxwell relation):
\begin{equation}
\dfrac{\partial \langle V \rangle}{\partial (\frac{1}{T}) } = \dfrac{\partial \langle U \rangle}{\partial (\frac{P}{T}) }
\label{eq:Maxwell_eq}
\end{equation}
the $\psi$ metric \eqref{eq:psi_deriv2} reduces to\footnote{The first variance is given erroneously, without the factor of 2, by Callen \cite{Callen_1960, Callen_1985}.}:  %Callen 1985, p426
\begin{align}
\frac{\GAMMA_{eq}}{k} 
=T\langle V \rangle \begin{bmatrix} \compress_{{T}} {P}^{2} - 2\, \alpha PT + {\frac {C_{{P}}T}{\langle V \rangle}}, \hspace{5pt} & \alpha T - \compress_{{T}}P \\
\alpha T - \compress_{{T}}P, &\compress_{{T}} \end{bmatrix} 
\label{eq:metric_psi}
\end{align}
whence from \eqref{eq:Legendre_eq}: 
\begin{align}
k \G_{eq}
= \frac{1}{{T}^{2} (\compress_{{T}}C_{{P}} - {\alpha}^{2}T\langle V \rangle)}
\begin {bmatrix} \compress_{{T}}, &\compress_{{T}}P - \alpha T
\\ \compress_{{T}}P - \alpha T, \hspace{5pt} &  \compress_{{T}}{P}^{2}
-2 \alpha PT + \frac{C_{{P}}T}{\langle V \rangle} \end {bmatrix}
\label{eq:metric_Sstar}
\end{align}
Using \eqref{eq:Sstar_deriv1}, \eqref{eq:psi_deriv1} and \eqref{eq:metric_psi}-\eqref{eq:metric_Sstar}, the (dimensional) arc lengths \eqref{eq:L2_Hstar}-\eqref{eq:L2_phi} and action integrals \eqref{eq:deltaH_tot_int}-\eqref{eq:deltaphi_tot_int} are obtained as:
\begin{gather}
\begin{split}
&\breve{L}_{S^*} 
= \int\limits_0^{\xi_{max}} \sqrt{ {\vect{\dot{f}}_{eq}}^{\top} \, k \G_{eq} \, \vect{\dot{f}}}_{eq} \, d\xi 
%\\&
= \int\limits_{0}^{\xi_{max}} \sqrt{ \frac{ \langle V \rangle \langle \dot{U} \rangle \bigl[\compress_T  \langle \dot{U} \rangle +  2 \langle \dot{V} \rangle (\compress_T P -  \alpha T) \bigr] + \langle \dot{V} \rangle^2 \bigl[P \langle V \rangle (\compress_T P - 2 \alpha T)  + C_P T \bigr]}{T^2 \langle V \rangle(C_P \compress_T - \alpha^2 T\langle V \rangle)}} d\xi
\end{split}
\label{eq:L_S} 
\\
\begin{split}
&\breve{\mathcal{J}}_{S^*} 
=  \int\limits_0^{\xi_{max}} \frac{1}{2} \; { {\vect{\dot{f}}_{eq}}^{\top} \, k \G_{eq} \, \vect{\dot{f}}}_{eq} \, d\xi 
%\\ &
= \int\limits_{0}^{\xi_{max}} {\frac{ \langle V \rangle \langle \dot{U} \rangle \bigl[\compress_T  \langle \dot{U} \rangle +  2 \langle \dot{V} \rangle (\compress_T P - \alpha T) \bigr] + \langle \dot{V} \rangle^2 \bigl[P \langle V \rangle (\compress_T P - 2 \alpha T)  + C_P T \bigr]}{2 T^2 \langle V \rangle(C_P \compress_T - \alpha^2 T\langle V \rangle)}} d\xi
\end{split}
\label{eq:J_S} 
\\
\begin{split}
{\breve{L}_{\psi}} 
= \int\limits_0^{\xi_{max}} \sqrt{ {\vect{\dot{\Lambda}}_{eq}}^{\top} \, \frac{\GAMMA_{eq}}{k} \, \vect{\dot{\Lambda}}_{eq}} \, d\xi 
= \int\limits_{0}^{\xi_{max}} \sqrt{\frac{C_P \dot{T}^2 - 2 \alpha \langle V \rangle T \dot{T} \dot{P} + \compress_T \langle V \rangle T \dot{P}^2}{k^2 T^2}} d\xi
\end{split}
\label{eq:L_psi} 
\\
\begin{split}
{\breve{\mathcal{J}}_{\psi}}
&= \int\limits_0^{\xi_{max}} \frac{1}{2} \; {{\vect{\dot{\Lambda}}_{eq}}^{\top} \, \frac{\GAMMA_{eq}}{k} \, \vect{\dot{\Lambda}}_{eq}} \, d\xi 
= \int\limits_{0}^{\xi_{max}} {\frac{C_P \dot{T}^2 - 2 \alpha \langle V \rangle T \dot{T} \dot{P} + \compress_T \langle V \rangle T \dot{P}^2}{2 k^2 T^2}} d\xi
\end{split}
\label{eq:J_psi} 
\end{gather}
Using \eqref{eq:C_P}-\eqref{eq:alpha}, these two sets of measures can be shown to be equivalent. The above equations must be integrated along the particular thermodynamic path followed by the process, as defined by the velocities $\{\dot{T}, \dot{P}\}$ or $\{ \langle \dot{U} \rangle, \langle \dot{V} \rangle \}$. For a process which follows a pre-determined path, e.g.\ an adiabatic, isothermal, isovolumetric or isopiezometric curve, this can be simplified by expressing the velocities (e.g.\ $\dot{P}$) as functions of one independent velocity (e.g.\ $\dot{T}$). 

To comment on units:\ if the above quantities were calculated using the ``pure'' metrics $\G_{eq}$ or $\GAMMA_{eq}$, in either case the line element $ds_n$, arc length $L_n$ and the term $\bar{\epsilon}_n \mathcal{J}_n$ would be dimensionless (whence the action is in reciprocal $\xi$ units). Use of the ``natural'' metric $k \G_{eq}$, as conducted here, gives the line element and arc length in $\sqrt{J K^{-1}}$ and the action in ${J K^{-1}} \xi^{-1}$, consistent with $\Delta S^*_{tot} = \bar{\epsilon}_{S^*} \mathcal{J}_{S^*}$ being in entropy units. In contrast, use of the ``natural'' metric ${\GAMMA_{eq}}/{k}$ gives the line element and arc length in $\sqrt{K J^{-1}}$ and the action in ${K J^{-1}} \xi^{-1}$.  The latter case can be rescued by use of a modified line element $d\breve{s}_{\psi}' = \sqrt{ k {\vect{\dot{\Lambda}}_{eq}}^{\top} \, \frac{\GAMMA_{eq}}{k} \, k \vect{\dot{\Lambda}}_{eq}} \, d\xi$ -- as suggested by \eqref{eq:Sstar_deriv1} and \eqref{eq:psi_deriv2} -- giving the line element and arc length in $\sqrt{J K^{-1}}$ and the action in ${J K^{-1}} \xi^{-1}$.  Thus in both the $S^*$ and $\psi$ representations, the least action bound \eqref{eq:leastaction2} can be used to determine the {\it minimum entropy cost} of a transition from one equilibrium position to another, along a specified path on the manifold of equilibrium positions.  As noted earlier, for slow processes and constant $\epsilon_n$, this is attained by a process which proceeds at a constant thermodynamic speed $\dot{s}$ \cite{Nulton_S_1988, Salamon_etal_1988, Andresen_G_1994, Diosi_etal_1996} (a more general result is available for rapid processes \cite{Spirkl_Ries_1995}). For variable $\epsilon_n$ and/or for stepwise phenomena, the process should be divided into individual steps placed at equal distances along the arc length traversed by the process, giving the so-called ``equal thermodynamics distance'' principle\index{equal thermodynamic distance principle} \cite{Salamon_N_1998, Schaller_etal_2001, Salamon_etal_2001}. Such considerations have been applied to the optimisation of a wide variety of engineering and industrial batch and flow processes, including engine cycles, heat engines and pumps, chemical reactors, distillation towers and many other systems. % \cite{**}. %(for a length list, see {\it http://www.fys.ku.dk/\~andresen/BAhome/publ.ftt.html}). 

A final important point is that the minimum path length (double minimisation) principle \eqref{eq:doubleleastaction} -- involving calculation of the geodesic -- has been applied to the analysis of equilibrium systems\index{geodesic!equilibrium systems}.  In early work, this bound was established by applying the calculus of variations directly to particular thermodynamic problems, without use of a metric \cite{Salamon_N_A_B_1980, Andresen_1983_book}. More recently, such lower bounds have been examined for particular thermodynamic systems \cite{Salamon_N_1998, Schon_1996, Brody_H_2009}.  In either case, for the entropy representation, this method yields an {\it absolute minimum entropy cost} $\Delta S^* \ge \Delta S^*_{min}$ for a transition between two equilibrium positions at particular rates of change, irrespective of the path. For cyclic or flow processes, this therefore gives a {\it minimum entropy production principle}\index{minimum entropy production principle!equilibrium systems} $\dot{S} \ge \dot{S}_{min}$, providing one of the key concepts of finite-time (or finite-parameter) thermodynamics. % \cite{**}. 

%%%%%%%%%%%%%%%%%%%%%%%%%%%%%%%%%%%
\subsection{\label{Flow}Flow Systems}
We now consider a flow system consisting of a control volume, subject to continuous flows of heat, particles and momentum, and within which chemical reactions may take place.  A few workers have examined such non-equilibrium systems previously within a Riemannian context, including for the Onsager linear regime \cite{Nathanson_S_1980, Gilmore_1982} and for extended irreversible thermodynamics \cite{Sien_B_1991, Chen_2001, Chen_2003}. A different perspective is provided here, based on a recent analysis of a flow system from a Jaynesian perspective\index{flow systems}\index{steady state systems}\index{least action bound!steady state systems} \cite{Niven_MEP}. This involves a probabilistic analysis of each infinitesimal element of the control volume, which experiences instantaneous values of the heat flux ${\vect j}_{Q,\flow{I}}$, mass fluxes ${\vect j}_{\flow{N}_c}$ of each species $c$, stress tensor ${{\tens{\tau}}}_{\flow{J}}$ and molar rate per unit volume $\hat{\dot{\xi}}_{\flow{L}_{d}}$ of each chemical reaction $d$, where the indices $\flow{I},\flow{J}, \flow{L}_{d}, \flow{N}_c \in \{0, \pm 1, \pm 2, ...\}$.  We therefore consider the joint probability $\pi_{\vecti} =  \pi_{\flow{I},\flow{J},\{\flow{L}_d\},\{\flow{N}_c\}}$ of instantaneous fluxes through the element and instantaneous reactions within the element, giving the (dimensionless) ``flux entropy'' function\index{flux entropy}:
\begin{align}
\mathfrak{H}_{st} &=   -  \sum\limits_{\vecti}  \pi_{\vecti} \ln \pi_{\vecti} ,
\label{eq:Hst}
\end{align}
Again assuming that each joint level $\vecti$ is equally probable, \eqref{eq:Hst} is maximised subject to constraints on the mean values of the heat flux $\langle {\vect j}_Q \rangle$, mass fluxes $\langle {\vect j}_{c} \rangle$, stress tensor $\langle {\tens{\tau}} \rangle$ and molar reaction rates $\langle \hat{\dot{\xi}}_d \rangle$ through or within the element, as well as by the natural constraint (\ref{eq:C0}). This gives the {\it steady state position} of the system:
\begin{gather}
\pi_{\vecti}^* = \frac{1}{\flow{Z}} \exp{  \biggl(  - \vect{\zeta}_{Q} \cdot {\vect j}_{Q,\flow{I}} - \sum\limits_{c} \vect{\zeta}_c \cdot {\vect j}_{ \flow{N}_c}} - {\tens \zeta}_{\tens \tau} : {\tens \tau}_{\flow{J}} 
 - \sum\limits_{d} \zeta_{d} \hat{\dot{\xi}}_{\flow{L}_d} \biggr)
\label{eq:pstar2st} 
\end{gather}
where $\vect{\zeta}_{Q}, \vect{\zeta}_c, {\tens \zeta}_{\tens \tau}$ and $\zeta_{d}$ are the Lagrangian multipliers associated with the heat, particle, momentum and chemical reaction constraints, and $\flow{Z} = e^{\zeta_0}$ is the partition function. By a traditional control volume analysis\index{control volume analysis} \cite{deGroot_M_1962, Prigogine_1967, Kreuzer_1981, Bird_etal_2006}, the multipliers can be identified as \cite{Niven_MEP}:
\begin{align}
{\vect{\zeta}}_Q &= - \frac {\tscale \Vscale}{k} {\vect \nabla} \biggl( {\frac{1}{T}} \biggr) 
  \label{eq:ass_zeta_Q}
\\
 {\vect{\zeta}}_c &= \frac {\tscale \Vscale}{k} \biggl[ {\vect \nabla} \biggl( \frac{\mu_c}{M_c T} \biggr) - \frac {{\vect F}_c}{T}   \biggr]
  \label{eq:ass_zeta_c}
\\
 {\tens{\zeta}}_{\tens{\tau}} &=  \frac {\tscale \Vscale}{k} { \vect \nabla} \biggl( \frac { \vect v }{T} \biggr)^\top
\label{eq:ass_zeta_Pi}
\\
 \zeta_d &= \frac {\tscale \Vscale}{k} \frac {A_d}{T}
 \label{eq:ass_zeta_d}
\end{align}
where %$T$ is absolute temperature, $P$ is absolute pressure, 
$\mu_c$ is the chemical potential of the $c$th constituent, $M_c$ is the molar mass of the $c$th constituent, $\vect{F}_c$ is the specific body force on species $c$, $\vect{v}$ is the mass-average velocity, $A_d$ is the chemical affinity of the $d$th reaction ($<0$ for a spontaneous reaction), $\vect{\nabla}$ is the Cartesian gradient operator, and $\tscale$ and $\Vscale$ respectively are characteristic time and volume scales of the system.  Generalising each component of the above multipliers as $\zeta_r$ and constraints as $\langle {j}_r \rangle$ with $r\in \{1,...,R\}$, Jaynes' relations \eqref{eq:Hstar}-\eqref{eq:Legendre} and \eqref{eq:Massieu} reduce to\index{manifold! of steady state positions}:
\begin{gather}
\mathfrak{H}^*_{st} = \ln \flow{Z}  + \sum_{r=1}^R \zeta_r \langle j_r \rangle = - \phi_{st}  - \frac{\theta \flow{V}}{k} \hat{\dot{\sigma}}
\label{eq:Hstar_st}
\\
\vect{\Lambda}_{st} = \biggl[ \frac{\partial \mathfrak{H}^*_{st}}{\partial \langle j_1 \rangle }, ..., \frac{\partial \mathfrak{H}^*_{st}}{\partial \langle j_R \rangle} \biggr]^{\top} = \bigl[ \zeta_1, ..., \zeta_R \bigr]^{\top}  
\label{eq:Hstar_deriv1_st}
\\
- \G_{st}  
= \begin{bmatrix} \dfrac{\partial^2 \mathfrak{H}^*_{st}}{\partial \langle j_1 \rangle^2} &... & \dfrac{\partial^2 \mathfrak{H}^*_{st}}{\partial \langle j_1 \rangle \partial \langle j_R \rangle} \\
\vdots &\ddots &\vdots\\
\dfrac{\partial^2 \mathfrak{H}^*_{st}}{\partial \langle j_R \rangle \partial \langle j_1 \rangle} &... &\dfrac{\partial^2 \mathfrak{H}^*_{st}}{\partial \langle j_R \rangle^2} \end{bmatrix} 
= \begin{bmatrix} \dfrac{\partial \zeta_1}{\partial \langle j_1 \rangle} &... & \dfrac{\partial \zeta_R}{\partial \langle j_1 \rangle } \\
\vdots &\ddots &\vdots\\
\dfrac{\partial \zeta_1}{\partial \langle j_R \rangle} &... &\dfrac{\partial \zeta_R}{\partial \langle j_R \rangle} \end{bmatrix} 
\label{eq:Hstar_deriv2_st}
\\
\phi_{st}  =  - \ln \flow{Z} =- \mathfrak{H}^*_{st}  +  \sum_{r=1}^R \zeta_r \langle j_r \rangle = - \mathfrak{H}^*_{st}
- \frac{\theta \flow{V}}{k} \hat{\dot{\sigma}}
\label{eq:phi_st}
\\
\vect{f}_{st} = 
\biggl[ \frac{\partial \phi_{st}}{\partial \zeta_1},..., \frac{\partial \phi_{st}}{\partial \zeta_R} \biggr]^{\top} = \bigl[ \langle j_1 \rangle,..., \langle j_R \rangle \bigr]^{\top}
\label{eq:phi_deriv1_st}
\\
- {\GAMMA_{st}}
= \begin{bmatrix} \dfrac{\partial^2 \phi_{st}}{\partial \zeta_1^2} &... & \dfrac{\partial^2 \phi_{st}}{\partial \zeta_1 \partial \zeta_R} \\
\vdots &\ddots &\vdots\\
\dfrac{\partial^2 \phi_{st}}{\partial \zeta_R \partial \zeta_1} &... &\dfrac{\partial^2 \phi_{st}}{\partial \zeta_R^2} \end{bmatrix} 
= \begin{bmatrix} \dfrac{\partial \langle j_1 \rangle}{\partial \zeta_1} &... & \dfrac{\partial \langle j_R \rangle }{\partial \zeta_1} \\
\vdots &\ddots &\vdots\\
\dfrac{\partial \langle j_1 \rangle}{\partial \zeta_R} &... &\dfrac{\partial \langle j_R \rangle}{\partial \zeta_R} \end{bmatrix} 
\label{eq:phi_deriv2_st}
\\
\G_{st} \; {\GAMMA_{st}} = \I
\label{eq:Legendre_st}
\end{gather}
where $\hat{\dot{\sigma}}$ can be identified as the local entropy production per unit volume (units of $J K^{-1} m^{-3} s^{-1}$). A flow system subject to constant flux and reaction rate constraints will therefore converge to a steady state position defined by a maximum in the flux entropy $\mathfrak{H}^*_{st}$ and a minimum in the flux potential $\phi_{st}$. If these effects occur simultaneously, the system will converge to a position of maximum $\hat{\dot{\sigma}}$, therefore providing a conditional, local derivation of the {\it maximum entropy production} (MEP) principle\index{maximum entropy production principle} \cite{Niven_MEP}, which has been applied as a discriminator to determine the steady state of many non-linear flow systems \cite{Paltridge_1975, Paltridge_1978, Paltridge_1981, Ozawa_etal_2003, Dewar_2003, Dewar_2005, Kleidon_L_book_2005, Martyushev_S_2006, Bruers_2007c}. 

In Onsager's analysis of transport phenomena in the vicinity of equilibrium \cite{Onsager_1931a, Onsager_1931b}, the fluxes and reaction rates are considered to be linear functions of the ``forces'' (the driving gradients and chemical affinities). In the present terminology, this would be written as:
\begin{equation}
\langle j_r \rangle %= \sum\limits_m L_{rm}^0 F_m 
= K \sum\limits_m L_{rm}^0 {\zeta_m}
\label{eq:Onsager}
\end{equation}
where $L_{rm}^0$ are the (constant) phenomenological coefficients at the zero-gradient position (i.e., at equilibrium) and $K=k/\theta \flow{V}$. In the present analysis, we do not claim linearity between $\langle j_r \rangle$ and $\zeta_m$, nor consider that the system is ``close to equilibrium'', but simply adopt the partial derivatives ${\partial \langle j_r \rangle}/{\partial \zeta_m}$ within the metric $\GAMMA_{st}$ \eqref{eq:phi_deriv2_st} as a set of parameters (functions of $\zeta_m$) with which to analyse the system. The present analysis therefore encompasses, but is not restricted to, Onsager's linear regime\index{Onsager linear regime}. The diagonal and many off-diagonal terms can readily be identified as functions of the conductivities (transport coefficients) and chemical reaction rate coefficents\index{susceptibilities!steady state systems} \cite{Bird_etal_2006}:
\begin{alignat}{2}
\hspace{5pt} &\text{Heat conductivity:} \hspace{10pt} &&\widetilde{\kappa}_{\imath \jmath} = - \frac{\partial \langle j_{Q\imath} \rangle}{\partial \biggl(\dfrac{\partial T}{\partial \jmath} \biggr)} 
\label{eq:heatcond_coeff}
\\
&\text{Diffusion coefficient, species }c\text{:} \hspace{10pt} &&\widetilde{D}^c_{\imath \jmath} = - \frac{\partial \langle j_{c \imath} \rangle}{\partial \biggl(\dfrac{\partial \hat{C}_c}{\partial \jmath} \biggr)} 
\label{eq:diffusion_coeff}
\\
&\text{Viscosity coefficient:} \hspace{10pt} &&\widetilde{\mu}_{\imath \jmath \kappa \ell} = - \frac{\partial \langle \tau_{\imath \jmath} \rangle}{\partial \biggl(\dfrac{\partial v_{\kappa}}{\partial \ell} \biggr)} 
\label{eq:viscosity}
\\
&\text{Rate coefficient, reaction }d\text{:} \hspace{10pt} &&\widetilde{k}_{d} = \frac{\partial \langle \hat{\dot{C}}_{cd} \rangle}{\partial \hat{C}_c} 
= \nu_{cd} M_c \frac{\partial \langle \hat{\dot{\xi}}_{d} \rangle}{\partial \hat{C}_c} 
\label{eq:reaction_rate_coeff}
\end{alignat}
where $\hat{C}_c$ is the concentration of species $c$ (units of kg m$^{-3}$; often used as a proxy for the chemical potential $\mu_c$), $\langle \hat{\dot{C}}_{cd} \rangle$ is the mean rate of change of concentration of species $c$ in the $d$th reaction (units of kg m$^{-3}$ s$^{-1}$), $\nu_{cd}$ is the stoichiometric coefficient of species $c$ in the $d$th reaction (positive if a product), and the indices $\imath, \jmath, \kappa, \ell \in \{x,y,z\}$. The remaining off-diagonal terms consist of the cross-process conductivity coupling coefficients and conductivity-reaction rate coefficients.  The Riemannian metric $\GAMMA_{st}$ can therefore be regarded as a function of the material properties or susceptibilities of a flow and chemical reactive system, in the same way that the Riemannian metric for an equilibrium system $\GAMMA_{eq}$ is a function of its various susceptibilities, such as $C_P$, $\compress_T$ and $\alpha$ (\S\ref{Equil}). As with equilibrium systems, an abrupt change in a given component $\gamma_{st,rm}$ with $\zeta_m$ can be interpreted as the boundary of a phase change in the system. Notice also that symmetry of $\GAMMA_{st}$ yields a set of Maxwell-like relations\index{Maxwell relations} \cite{Niven_MEP}:
\begin{equation}
\frac{\partial \langle j_r \rangle}{\partial \zeta_m} = \frac{\partial \langle j_m \rangle}{\partial \zeta_r}
\label{eq:Maxwell_st}
\end{equation}
These apply to all infinitesimal volume elements of a flow system, not merely those in the vicinity of equilibrium. Eqs.\ \eqref{eq:Maxwell_st} considerably simplify the set of parameters needed for analysis, from $R^2$ to $\bigl( \begin{smallmatrix} R+1\\ 2 \end{smallmatrix} \bigr)$ coefficients; further simplifications may  be attainable in certain systems due to geometric and tensor symmetries \cite{Bird_etal_2006}.

The above relations \eqref{eq:Hstar_st}-\eqref{eq:phi_deriv2_st} can now be applied to develop a Riemannian description of a flow system on the manifold of steady state positions.  In terms of the generalised derivatives, the (dimensionless) arc lengths \eqref{eq:L2_Hstar}-\eqref{eq:L2_phi} and action integrals \eqref{eq:deltaH_tot_int}-\eqref{eq:deltaphi_tot_int} are obtained as:
\begin{gather}
\begin{split}
{L}_{st} &= \int\limits_0^{\xi_{max}} \sqrt{{\vect{\dot{f}}_{st}}^{\top} \, \G_{st} \, \vect{\dot{f}}}_{st} \, d\xi 
= \int\limits_0^{\xi_{max}} \sqrt{{\vect{\dot{\Lambda}}_{st}}^{\top} \, {\GAMMA_{st}} \, \vect{\dot{\Lambda}}_{st}} \, d\xi 
%\\&
= \int\limits_0^{\xi_{max}} \sqrt{- \vect{\dot{\Lambda}}_{st} \cdot {\vect{\dot{f}}_{st}}}  \; d\xi
= \int\limits_0^{\xi_{max}} \sqrt{- \sum\limits_{r=1}^R \frac{\partial \zeta_r}{\partial \xi} \frac{\partial \langle j_r \rangle}{\partial \xi}} \; d\xi
\end{split}
\label{eq:L_st} 
\\
\begin{split}
{\mathcal{J}}_{st} 
&=  \int\limits_0^{\xi_{max}} \frac{1}{2} \; { {\vect{\dot{f}}_{st}}^{\top} \, \G_{st} \, \vect{\dot{f}}}_{st} \, d\xi 
= \int\limits_0^{\xi_{max}} \frac{1}{2} \; {{\vect{\dot{\Lambda}}_{st}}^{\top} \, {\GAMMA_{st}} \, \vect{\dot{\Lambda}}_{st}} \, d\xi
%\\ &
= - \int\limits_0^{\xi_{max}} \frac{1}{2}{\vect{\dot{\Lambda}}_{st} \cdot {\vect{\dot{f}}_{st}}}  \; d\xi
= - \frac{1}{2} \int\limits_0^{\xi_{max}} \sum\limits_{r=1}^R {\frac{\partial \zeta_r}{\partial \xi} \frac{\partial \langle j_r \rangle}{\partial \xi}} \; d\xi
\end{split}
\label{eq:J_st} 
\end{gather}
where, as shown, the two alternative $\mathfrak{H}^*_{st}$ and $\phi_{st}$ measures are equivalent. Once again, these equations must be integrated along the particular path taken between the initial and final steady state positions. 

To comment on units:\ since the above quantities are calculated using the ``pure'' metrics $\G_{st}$ or $\GAMMA_{st}$, the resulting line element $ds_{st}$, arc length $L_{st}$ and the term $\bar{\epsilon}_{st} \mathcal{J}_{st}$ are dimensionless. Use of the ``natural'' metric $K \G_{st}$, for $K=k/\theta \flow{V}$, therefore gives the line element and arc length in $\sqrt{J K^{-1} m^{-3} s^{-1}}$ and the action in ${J K^{-1} m^{-3} s^{-1}} \xi^{-1}$, thereby giving $\bar{\epsilon}_{st} \mathcal{J}_{st}$ in units of entropy production per unit volume. Similarly, use of the ``natural'' metric ${\GAMMA_{st}}/{K}$ in conjunction with the dimensional constraint vector $K {\vect{\dot{\Lambda}}_{st}}$ gives the line element and arc length in $\sqrt{J K^{-1} m^{-3} s^{-1}}$ and action in ${J K^{-1} m^{-3} s^{-1}} \xi^{-1}$, again giving $\bar{\epsilon}_{st} \mathcal{J}_{st}$ in units of entropy production per unit volume.  
The least action bound \eqref{eq:leastaction2} therefore yields a {\it minimum entropy production principle}\index{minimum entropy production principle!steady state systems}, which sets a lower bound for the entropy production associated with movement of a flow system from one steady state position to another along a specified path. From the previous analysis, this involves two separate minimisation principles:
\begin{list}{$\bullet$}{\topsep 3pt \itemsep 3pt \parsep 0pt \leftmargin 8pt \rightmargin 0pt \listparindent 0pt
\itemindent 0pt}
\item If the path is specified, the process of minimum entropy production will be one which proceeds at constant speed $\dot{s}$, assuming a slow process and a constant dissipation parameter $\epsilon$. Alternately, if the dissipation parameter $\epsilon$ is not constant, the minimum entropy production process will be given by a constant arc length speed, in accordance with a steady state analogue of the ``equal thermodynamic distance'' principle \cite{Nulton_etal_1985, Salamon_N_1998, Schaller_etal_2001, Salamon_etal_2001}. 
\item If the path is not specified or can be varied, an absolute lower bound for the entropy production is given by the geodesic in steady state parameter space\index{geodesic!steady state systems}, in accordance with the methods of \S\ref{Geodesic}.
\end{list}
Although they share a similar name, the minimum entropy production  principle derived herein is quite different to that of Prigogine \cite{Prigogine_1967}, which concerns the selection of a steady state position relative to possible non-steady state positions, and which only applies to the Onsager linear regime. Similarly, it differs from the minimum entropy production principle obtained by the application of Riemannian geodesic calculations to the manifold of equilibrium positions, discussed at the end of \S\ref{Equil} \cite{Salamon_N_A_B_1980, Andresen_1983_book, Salamon_N_1998, Schon_1996, Brody_H_2009}.  The minimum principle derived herein is more general than both these principles, being applicable beyond the set of equilibrium positions, and also well outside the linear regime of non-equilibrium thermodynamics.  In turn, it is based on the even broader generic formulation of the least action bound given herein, applicable to any system which can be analysed by Jaynes' method.

%%############################################################################
\section{\label{sect:Concl}Conclusions}
%% ############################################################################
In this study, the manifold of stationary positions inferred by Jaynes' MaxEnt and MaxREnt principles  -- considered as a function of the moment constraints or their conjugate Lagrangian multipliers -- is endowed with a Riemannian geometric description, based on the second differential tensor of the entropy or its Legendre transform (negative Massieu function) obtained from Jaynes' method. The analysis provides a generalised {\it least action bound} applicable to all Jaynesian systems, which provides a lower bound to the cost (in generic entropy units) of a transition between inferred positions along a specified path, at specified rates of change of the control parameters. The analysis therefore extends the concepts of ``finite time thermodynamics'', developed over the past three decades \cite{Weinhold_1975a, Weinhold_1975b, Weinhold_1975c, Weinhold_1975d, Weinhold_1976, Salamon_A_G_B_1980, Salamon_N_A_B_1980, Salamon_B_1983, Salamon_I_B_1983, Andresen_1983_book, Salamon_N_I_1984, Andresen_B_O_S_1984, Salamon_N_B_1985, Schlogl_1985, Feldman_etal_1985, Nulton_etal_1985, Feldman_etal_1986, Levine_1986, Nulton_S_1988, Salamon_etal_1988, Andresen_etal_1988a, Andresen_etal_1988b, Hoffmann_etal_1989, Andresen_G_1994, Salamon_N_1998, Salamon_etal_2001, Schaller_etal_2001, Beretta_1986, Beretta_1987, Beretta_2008, Diosi_etal_1996, Crooks_2007, Feng_C_2009, Brody_H_2009}, to the generic Jaynes domain, providing a link between purely static (stationary) inferred positions of a system, and dynamic transitions between these positions (as a function of time or some other coordinate).  If the path is unspecified, the analysis gives an absolute lower bound for the cost of the transition, corresponding to the geodesic of the Riemannian hypersurface.   

The analysis is then applied to (i) an equilibrium thermodynamic system subject to mean internal energy and volume constraints, and (ii) a flow system at steady state, subject to constraints on the mean heat, mass and momentum fluxes and chemical reaction rates.  The first example recovers the {\it minimum entropy cost} of a transition between equilibrium positions, a widely used result of finite-time thermodynamics. The second example leads to a new {\it minimum entropy production principle}, for the cost of a transition between steady state positions of a flow system. The analyses reveal the tremendous utility of Jaynes' MaxEnt and MinXEnt methods augmented by the generalised least action bound, for the analysis of probabilistic systems of all kinds.

%%%%%%%%%%%%%%%%%%%%%

%\ack
%\begin{acknowledgments}
\section*{Acknowledgments}
The first author thanks the European Commission for support as a Marie Curie Incoming International Fellow (FP6); The University of New South Wales, the University of Copenhagen and Technical University of Berlin for financial support; and Bob Dewar and Roderick Dewar for the opportunity to present this analysis at the 22nd Canberra International Physics Summer School, ANU, Canberra, December 2008.

%\end{acknowledgments}

%############################################################################
%\begin{appendix}
\appendix
\section{\label{Apx1}Riemannian Geometric Considerations}

It is necessary to examine several salient features of the Riemannian geometric interpretation adopted herein \cite{Kreysig_1991, Zwillinger_2003}. Consider a hypersurface represented by the position vector $\vect{x}=[x_1,...,x_n]^{\top}$, embedded within the $n$-dimensional space defined by the coordinates $(x_1,...,x_n)$. For analysis, this hypersurface can be converted to the parametric representation $\vect{x}(\vect{u})=[x_1(\vect{u}),...,x_n(\vect{u})]^{\top}$, where $\vect{u}=[u_1,...,u_{n-1}]^{\top}$ is the $(n-1)$-dimensional vector of parameters $u_j$, consisting of  coordinates on the hypersurface. The {\it first fundamental form} of this geometry is defined by the metric \cite{Kreysig_1991, Zwillinger_2003}:
\begin{equation}
d\varsigma^2 = d\vect{x} \cdot d\vect{x} = \sum\limits_{i=1}^{n-1} \sum\limits_{j=1}^{n-1} a_{ij} du_i du_j = d\vect{u}^{\top} \, \tens{a} \, d\vect{u}
\label{eq:fff}
\end{equation}
in which, by elementary calculus, the components of the tensor $\tens{a}$ can be shown to be:
\begin{equation}
a_{ij} = \frac{\partial \vect{x}}{\partial u_i} \cdot \frac{\partial \vect{x}}{\partial u_j}
\label{eq:fff_tens}
\end{equation}
Accordingly, $\tens{a}$ is symmetric. By Euclidean geometry, \eqref{eq:fff} can be used to calculate distances between two points $a$ and $b$ on the hypersurface $\vect{x}$, on the path defined by $\vect{u}$:
\begin{equation}
L_{\vect{x}} = \int_a^b d\varsigma = \int_a^b \sqrt{d\vect{u}^{\top} \, \tens{a} \, d\vect{u}} = \int_{\xi_a}^{\xi_b} \sqrt{d\vect{\dot{u}}^{\top} \, \tens{a} \, d\vect{\dot{u}}} \; d\xi
\label{eq:fff_L}
\end{equation}
where the overdot indicates the derivative with respect to the path parameter $\xi$.  The {\it second fundamental form} of the hypersurface is then defined by \cite{Kreysig_1991, Zwillinger_2003}:
\begin{equation}
- d\vect{x} \cdot d\vect{n} = \sum\limits_{i=1}^{n-1} \sum\limits_{j=1}^{n-1} b_{ij} du_i du_j = d\vect{u}^{\top} \, \tens{b} \, d\vect{u}
\label{eq:sff}
\end{equation}
where $\vect{n}$ is the unit normal vector to the hypersurface.  By differential calculus, it can be shown that:
\begin{gather}
%\vect{n} = ***
%\label{eq:nvect} 
%\\
b_{ij} = \frac{\partial \vect{x}}{\partial u_i \partial u_j} \cdot \vect{n}
\label{eq:sff_tens} 
\end{gather}
The second fundamental form is not considered as a metric with which to calculate distances, but is used to examine the tangency and curvature properties of the manifold $\vect{x}$ \cite{Kreysig_1991, Zwillinger_2003}. 

In the present study, we wish to adopt the Jaynesian matrix $\G$ or $\GAMMA$ as a Riemannian metric tensor for the calculation of arc lengths on the $R$-dimensional stationary state hypersurface, embedded in the $(R+1)$-dimensional space defined by $(\mathfrak{H}^*, \{ \langle f_r \rangle \})$ or $(\phi, \{\lambda_r \})$. We therefore adopt the (somewhat peculiar) approach in which the coordinates $[x_2, ..., x_{R+1}]^{\top}$ are selected as the surface parameters $[u_1, ..., u_R]^{\top}$; i.e.\ with the hypersurface $\vect{x}_{\mathfrak{H}^*} = [\mathfrak{H}^*, \langle f_1 \rangle, ..., \langle f_R \rangle]^{\top}$ parameterised by $\vect{u}_{\mathfrak{H}^*} = \vect{f}$ and with $\vect{x}_{\phi} = [\phi, \lambda_1, ..., \lambda_R]^{\top}$ parameterised by $\vect{u}_{\phi}=\vect{\Lambda}$. Two necessary conditions for the use of $\G$ or $\GAMMA$ as metric tensors is that they be symmetric and positive definite (or semi-definite); since they constitute Hessian matrices of the concave generic entropy $\mathfrak{H}^*$ or convex potential function $\phi$, these conditions are satisfied, not only in thermodynamic applications but within the generic Jaynes formulation (with semi-definite behaviour only at singularities) \cite{Salamon_A_G_B_1980, Kapur_K_1992}. However, $\G$ and $\GAMMA$ are related to a second, rather than a first, fundamental form \cite{Salamon_A_G_B_1980, Andresen_etal_1988b}. For $\G$ or $\GAMMA$ to be considered as metric tensors, they must be able to generate the first fundamental form of some position vector which describes the hypersurface. In mathematical terms, from \eqref{eq:fff}:
\begin{align}
ds_{\mathfrak{H}^*}^2 
&= {d \vect{f}^{\top} \, \G \, d \vect{f}}  =d {\vect{u}_{\mathfrak{H}^*}}^{\top} \, \tens{a}_{\mathfrak{H}^*} \, d\vect{u}_{\mathfrak{H}^*}, 
\label{eq:ds_Hstar_fff} \\
ds_{\phi}^2 
&= {d \vect{\Lambda}^{\top} \, \GAMMA \, d \vect{\Lambda}} = d{\vect{u}_{\phi}}^{\top} \, \tens{a}_{\phi} \, d\vect{u}_{\phi}
\label{eq:ds_phi_fff} 
\end{align}
From \eqref{eq:MaxREnt_deriv2}, \eqref{eq:Massieu_deriv2}, \eqref{eq:Massieu} and \eqref{eq:fff_tens}, taking advantage of tensor symmetries, the metric components must therefore satisfy:
\begin{align}
- g_{mr} &= - a_{{\mathfrak{H}^*}, mr} 
= \frac{\partial \vect{\omega}}{\partial \langle f_m \rangle} \cdot \frac{\partial \vect{\omega}}{\partial \langle f_r \rangle} 
= \frac{\partial ^2 \mathfrak{H}^*}{\partial \langle f_m \rangle \partial \langle f_r \rangle} 
= \frac{{\partial \lambda_r }}{{\partial \langle {f_m } \rangle }}
\label{eq:ds_Hstar_fff2} \\
- \gamma_{mr} &= - a_{\phi, mr} 
= \frac{\partial \vect{\Omega}}{\partial \lambda_m} \cdot \frac{\partial \vect{\Omega}}{\partial \lambda_r}
= \frac{\partial^2 \phi}{\partial \lambda_m \partial \lambda_r}
= \frac {\partial \langle {f_r} \rangle}{\partial \lambda_m}  
\label{eq:ds_phi_fff2} 
\end{align}
where $\vect{\omega}(\vect{f})$ and $\vect{\Omega}(\vect{\Lambda})$ are new $R$-dimensional position vectors, which from \eqref{eq:Legendre2}, are related by:
\begin{equation}
\tens{a}_{\mathfrak{H}^*} \, \tens{a}_{\phi} = \I,
\label{eq:Legendre2}
\end{equation}
In consequence, the metrics \eqref{eq:ds_Hstar}-\eqref{eq:ds_phi} and \eqref{eq:ds_Hstar2}-\eqref{eq:ds_phi2} and arc lengths \eqref{eq:L1_Hstar}-\eqref{eq:L2_phi} used herein are not measures of distance on the stationary state hypersurface defined by $\mathfrak{H}^* (\{ \langle f_r \rangle \})$ or $\phi(\{\lambda_r \})$, but rather, on the transformed hypersurface given by $\vect{\omega}$ or $\vect{\Omega}$. In addition to the symmetry and positive definiteness conditions, it is therefore also necessary and sufficient that the hypersurface defined by $\vect{\omega}$ or $\vect{\Omega}$ exists within $\mathbb{R}^R$, is continuous and continuously differentiable -- at least up to first order -- except in the neighbourhood of singularities.

%\end{appendix}

%\begin{appendix}
%\section*{\label{Apx2}Appendix 2}
%
%\end{appendix}
%% ############################################################################

%% ############################################################################
%\section*{References}

%\bibliographystyle{ws-rv-van}
%\bibliography{ws-rv-sample}
%\printindex[aindx]
%\printindex

\end{document}